%% file: main.tex
\documentclass[%
 reprint,
 superscriptaddress,
 amsmath,amssymb,
 aps,
 prb,
]{revtex4-2}
\usepackage{url}
\usepackage{hyperref}
\usepackage{xcolor}
\usepackage{graphicx}
\graphicspath{{fig/BFS}{fig/COND}{fig/DOS}{fig/Fig}}
\usepackage{amsmath,mathrsfs}
\usepackage{dsfont}
\hypersetup{
	setpagesize=false,
	bookmarksnumbered=true,%
	bookmarksopen=true,%
	colorlinks=true,%
	linkcolor=magenta,
	citecolor=magenta,
}
\usepackage{dcolumn}
\usepackage{bm}
\usepackage{braket}
\usepackage{mhchem}
\usepackage{tikz}
\usetikzlibrary{intersections,arrows} 
\usepackage{ulem}
\usepackage{silence}
\usepackage[whole]{bxcjkjatype}

\def\kk{{\bm{k}}}
\def\dd{{\bm{d}}}
\def\JJ{{\bm{J}}}
\DeclareMathOperator{\Ch}{Ch}
\DeclareMathOperator{\sign}{sign}
\DeclareMathOperator{\Diag}{Diag}
\DeclareMathOperator{\Tr}{Tr}
\DeclareMathOperator{\Pf}{Pf}
\newcommand{\mathbbm}[1]{\mathds{#1}}

\DeclareRobustCommand{\ykerase}{\bgroup\markoverwith{\textcolor{blue}{\rule[.5ex]{2pt}{0.4pt}}}\ULon}

\newcommand{\red}[1]{#1}

\begin{document}
\preprint{APS/123-QED}

\title{Surface density of states and tunneling spectroscopy of a spin-3/2 superconductor
with Bogoliubov Fermi Surfaces}

\author{Ryoi Ohashi}
\affiliation{Department of Materials Engineering Science, Osaka University, Toyonaka, Osaka 560-8531, Japan}

\author{Shingo Kobayashi}
\affiliation{RIKEN Center for Emergent Matter Science, Wako, Saitama 351-0198, Japan}

\author{Shotaro Kanazawa}
\affiliation{Department of Applied Physics, Nagoya University, Nagoya 464-8603, Japan}

\author{Yukio Tanaka}
\affiliation{Department of Applied Physics, Nagoya University, Nagoya 464-8603, Japan}
\affiliation{Research Center for Crystalline Materials Engineering, Nagoya University, Nagoya 464-8603, Japan}

\author{Yuki Kawaguchi}
\affiliation{Department of Applied Physics, Nagoya University, Nagoya 464-8603, Japan}
\affiliation{Research Center for Crystalline Materials Engineering, Nagoya University, Nagoya 464-8603, Japan}

\date{\today}

\begin{abstract}
Bogoliubov Fermi surfaces (BFSs) of superconducting states arise from point or line nodes by breaking time-reversal symmetry.
Because line and point nodes often accompany topologically protected zero-energy surface Andreev bound states (SABSs) and thereby lead to a characteristic zero-bias conductance peak (ZBCP) in tunneling spectroscopy, we investigate how these properties change when the line and point nodes deform into BFSs.
In this paper, we consider spin-quintet $J_{\rm pair}=2$ pairing states of spin-3/2 electrons with BFSs and calculate the surface density of states and the charge conductance. Comparing the obtained results with the cases of spin-singlet \textit{d}-wave pairing states having the same symmetry, we find that the ZBCP associated with point and/or line nodes is blunted or split in accordance with the appearance of the BFSs.
On the other hand, when the spin-singlet \textit{d}-wave state has point nodes but does not have SABS on the surface, we obtain a nonzero small electron conductivity at zero bias through the zero-energy states on the BFSs.
\end{abstract}
\maketitle

\section{Introduction}
 Nodal structures of a superconducting pair potential lead to the characteristic properties of unconventional superconductors. An example includes the power-law temperature dependence of bulk physical quantities at a low temperature, such as specific heat, spin relaxation rate, thermal conductivity, and London penetration depth~\cite{SigristUeda1991}. These signatures have been indeed observed in high-temperature superconductors and heavy fermion superconductors~\cite{Tsuei_2000,Stewart}.
Another characteristic originating from nodal structures is the appearance of surface Andreev bound states (SABSs) at zero energy due to the change of the internal phase degrees of freedoms of pair potentials~\cite{ABS,ABSb,Hu94,MS99,FMS01,VVE08}.  
The existence of zero-energy SABSs is related to the bulk topological properties of superconductors~\cite{SRFL08,qi11,tanaka2012,alicea12,Beenakker13,SatoAndo2017,tanaka_2024_Progress}. Namely, there is a one-to-one correspondence between the number of zero-energy SABSs and the bulk topological invariant via the bulk-boundary correspondence. When the superconducting pair potential has line or point nodes, a topological invariant takes a nonzero value in a subspace of the Brillouin zone, leading to SABSs with dispersion or zero-energy flat-band SABSs in the surface Brillouin zone~\cite{Sato2011,Schnyder2011,Brydon2011,matsuura2013,Kobayashi14,Kobayashi16,Kobayashi18}.

Experimentally, SABSs manifest themselves in the low voltage profile of the tunneling spectroscopy of quasiparticles. 
For instance, in the case of spin-singlet \textit{d}-wave superconductors with line nodes, the zero-energy flat-band SABSs lead to a zero bias conductance peak (ZBCP) in normal metal/superconductor (N/S) junctions~\cite{TK95,KT96,Kashiwaya2000}, which was observed in tunneling spectroscopy experiments on high $T_{\rm c}$ cuprate superconductors~\cite{Alff97,Wei98,Iguchi2000,Kashiwaya2000}.
On the other hand, in the case of 3D chiral superconductors, which have a pair of point nodes in addition to a line node~\cite{Kobayashi2015,Tamura2017}, the dispersive SABSs originated from the point nodes also lead to a ZBCP, but the height of the ZBCP is strongly suppressed as compared to the case of the zero-energy flat-band SABSs~\cite{Yamashiro97,YTK98,TYBN09,TYN09,TanakaYokoyama2010,Yada2011,yamakage12,LuBo}.

Recently, a new type of nodal structure, called the Bogoliubov Fermi surface (BFS), is predicted for time-reversal symmetry (TRS) breaking superconductivity in a multi-band system~\cite{Agterberg2017, Brydon2018,Setty2020Nature,Setty2020,Timm2021,Kobayashi2022}, where zero-energy quasiparticle states appear on a surface in the 3D bulk Brillouin zone. 
The unique feature of this superconducting state is that the BFSs arise from the inflation of point or line nodes due to the splitting of superconducting energy bands around the nodes enforced by a TRS breaking effect~\cite{Brydon2018,Timm2017}. 
This implies that the BFSs are topological objects characterized by multiple topological invariants~\cite{Bzdusek2017}; namely, one characterizes the BFS and the others are associated with the underlying point or line nodes. 
It is expected that the multiple topological natures come together in tunneling spectroscopy and bring about a unique low voltage profile distinguished from the conventional ZBCP attributed to line or point nodes. 
Although tunneling spectroscopy in a superconductor with the BFSs has been studied in Refs.~\cite{Lapp2020,Banerjee2022,Setty2020,Dutta2023}, the influence of multiple topological natures on tunneling conductance has yet to be clarified. 

In addition, recent experiments on the iron-based superconductor \ce{Fe(Se,S)} have indicated the existence of the Fermi surface in the superconducting states. 
In these experiments, the signature of the BFS has been detected as a nonzero residual density of states at zero energy at a low temperature in specific heat, thermal conductivity, and tunneling spectroscopy~\cite{Sato2018Abrupt,Hanaguri2018,Mizukami2023unusual}.
More recently, the direct evidence of the BFS has been reported as an unusual superconducting gap anisotropy with zero gap regions using high-energy resolution laser-based angle-resolved photoemission spectroscopy~\cite{Nagashima2022discovery}. 
Thus, a detailed study on the influence of the BFSs on experimental observables is called for.

In this paper, we study the SABSs and tunneling spectroscopy of spin-quintet $J_\textrm{pair}=2$ pairing states of spin-3/2 electrons.
We calculate the surface density of states (SDOS) and the charge conductance in N/S junctions utilizing the recursive Green's function method~\cite{Umerskii} and the generalized Lee-Fisher formula~\cite{Lee1981,Inoue_text} which are available for the lattice models of Bogoliubov-de Gennes (BdG) Hamiltonian~\cite{KawaiYada,YadaGolubov,TamuraTakagi}. 
Here, we consider the three $J_\textrm{pair}=2$ states, $T_{2g}(0,i,1)$, $T_{2g}(1,\omega,\omega^2)$ and $E_{g}(1,i)$, all of which possess the BFSs. 
Because these states can be mapped to spin-singlet \textit{d}-wave pairing states with line and/or point nodes,
we compare the obtained results for the $J_\textrm{pair}=2$ pairing states with those for the corresponding spin-singlet \textit{d}-wave pairing states of spin-1/2 electrons.
The effects of the BFSs on the SABSs and the charge conductance are summarized as follows.
When the spin-singlet \textit{d}-wave state exhibits the zero-energy peak (ZEP) in the SDOS and the ZBCP due to the existence of the topologically protected zero-energy surface flat bands or surface arc states, the corresponding $J_\textrm{pair}=2$ pairing state has blunted or split peaks.
This is because the broken TRS in the $J_\textrm{pair}=2$ states violates the topological protection for the zero-energy SABSs in the \textit{d}-wave pairing states and makes the SABSs more dispersive.
On the other hand, in the case when the \textit{d}-wave state has no SABS, and hence, has zero SDOS and zero charge conductance at zero energy, the small SDOS at zero energy coming from the bulk BFSs and the charge conductance through the zero-energy states become visible in the $J_\textrm{pair}=2$ pairing state.
These features will serve as a guide to detect the BFSs by tunneling spectroscopy. 

The organization of this paper is as follows. 
In Sec.~\ref{sec:BFS}, we review how the BFSs appear in a doubly-degenerate two-band system.
In Sec.~\ref{sec:luttinger_kohn}, we introduce the Luttinger-Kohn model of spin-$3/2$ electrons with cubic symmetry and describe the pair potentials for spin-singlet $J_\textrm{pair}=0$ and spin-quintet $J_\textrm{pair}=2$ pairing states. 
Projection to the spin-singlet \textit{d}-wave pairing states of spin-1/2 electrons is also discussed here.
In Sec.~\ref{sec:setup}, we explain the detailed setup for our calculation and the calculation method.
In Sec.~\ref{sec:results}, we discuss the behavior of the SABSs in the presence of the BFSs with showing the numerical results of the SDOS and the charge conductance for $T_{2g}(1,i,0)$, $T_{2g}(1,\omega,\omega^{2})$, and $E_{g}(1,i)$ states.
In Sec.~\ref{sec:conclusion}, we conclude our results.

\section{Bogoliubov Fermi Surface in a doubly-degenerate two-band model}
\label{sec:BFS}
We briefly review the underlying physics of the appearance of BFSs~\cite{Brydon2018}.
BFSs can appear in even-parity superconducting states of inversion symmetric systems with broken TRS.
Examples include spin-singlet superconductors under an external magnetic field~\cite{Yang1998,Setty2020,zhu2020} and multi-band superconductors with TRS-breaking inter-band pairings. Here, we consider the latter case, i.e., the case when TRS is preserved in a normal state and spontaneously broken in a superconducting state.
A minimal model for such systems is a doubly-degenerate two-band model.

We start from a normal-state Hamiltonian given in the band basis as
\begin{align}
    \mathcal{H}_{\rm N}&=\sum_{\kk}(\bm c_\kk^\dagger)^{\rm T}H_0(\kk) \bm c_\kk,\\
    H_0(\kk)&=\Diag[E_{\kk,+},E_{\kk,+},E_{\kk,-},E_{\kk,-}]-\mu\mathbbm{1}_4,
\end{align}
where $E_{\kk,+}$ and $E_{\kk,-}$ are the energy dispersion of the two bands, each of which is doubly degenerate due to the TRS, $\mu$ is the chemical potential, $\mathbbm{1}_n$ denotes the $n\times n$ identity matrix, and
$\bm c_\kk=(c_{\kk+\uparrow},c_{\kk+\downarrow},c_{\kk-\uparrow}, c_{\kk-\downarrow})^{\rm T}$ and
$\bm c^\dagger_\kk=(c^\dagger_{\kk+\uparrow},c^\dagger_{\kk+\downarrow},c^\dagger_{\kk-\uparrow}, c^\dagger_{\kk-\downarrow})^{\rm T}$
with $c_{\kk\pm s}$ being the annihilation operator of an electron in the Bloch state $|\kk,\pm,s\rangle$ in the $\pm$ band with momentum $\kk$ and pseudospin $s=\uparrow,\downarrow$. 
From the assumptions of inversion symmetry and TRS,
the space inversion operator $P$ and the time-reversal operator $T$ transforms the Bloch states as
\begin{align}
    P|\kk,\pm,s\rangle&=|-\kk,\pm,s\rangle,\\
    T|\kk,\pm,s\rangle&=-\sign(s)|-\kk,\pm,-s\rangle,
\end{align}
where $\sign(\uparrow)=1$ and $\sign(\downarrow)=-1$. Due to the inversion symmetry, the energy dispersion satisfies $E_{\kk,\pm}=E_{-\kk,\pm}$.

The superconducting state is described by the Bogoliubov-de Gennes (BdG) Hamiltonian given by
\begin{align}
    \mathcal{H}&=\frac{1}{2}\sum_{\kk} \begin{pmatrix} (\bm c_\kk^\dagger)^{\rm T} & \bm c_{-\kk}^{\rm T}\end{pmatrix} 
    H_\kk
    \begin{pmatrix}\bm c_\kk \\ \bm c_{-\kk}^\dagger\end{pmatrix},\\
    H_\kk&=\begin{pmatrix}
    H_0(\kk) & \Delta(\kk) \\ \Delta^\dagger(\kk) & -H_0^{\rm T}(-\kk)
    \end{pmatrix},
\label{eq:H_BdG}
\end{align}
where the pairing term $\Delta(\kk)$ should have
even or odd parity.
For the case of even-parity pairing, which is our interest,
the general form of $\Delta(\kk)$ is given by
\begin{align}
    \Delta(\kk)=\begin{pmatrix}
    \psi_{\kk,+}i\sigma_y & (\psi_{\kk,I}\sigma_0+\bm d_\kk\cdot\bm \sigma)i\sigma_y\\
    (\psi_{\kk,I}\sigma_0-\bm d_\kk\cdot\bm \sigma)i\sigma_y & \psi_{\kk,-}i\sigma_y 
    \end{pmatrix},
    \label{eq:Delta_2band}
\end{align}
where $\bm \sigma=(\sigma_x,\sigma_y,\sigma_z)$ and $\sigma_0$ are the vector of Pauli matrices and the identity matrix in the pseudospin space, respectively,
and all functions in the matrix are even functions of $\bm k$.
Here, $\psi_{\kk,\pm}$ denotes the pseudospin-singlet pair in each band, whereas 
$\psi_{\kk,I}$ and $\dd_\kk$ are the pseudospin-singlet and triplet components, respectively, in the inter-band pairing.
Note that the pseudospin-triplet pairing should be band singlet due to the Fermi-Dirac statistics, which is the reason for the different signs in front of $\dd_\kk$ in the off-diagonal terms.
By defining the pseudospin operator for $+$ and $-$ bands as
\begin{align}
    \bm S_+=\begin{pmatrix} \bm \sigma & 0 \\ 0 & 0 \end{pmatrix},\ \ 
    \bm S_-=\begin{pmatrix} 0 & 0 \\ 0 & \bm \sigma \end{pmatrix},
\end{align}
the polarization in the pairing state at momentum $\kk$ is given by
\begin{align}
  \Tr[\Delta^\dagger (\kk) \bm S_\pm \Delta(\kk)] =2i \dd_\kk\times \dd_\kk^* \pm 4{\rm Re}(\psi_{\kk,I}^* \dd_\kk),
\label{eq:spin_polarization}
\end{align}
which means that TRS is spontaneously broken when $\dd_\kk$ is a complex vector such that $\dd_\kk^*\neq \dd_\kk$ or when the pseudospin-singlet and triplet components are mixed in the inter-band pairing.

BFS is a topological object defined for an inversion-symmetric BdG Hamiltonian with even-parity pairing.
Here, we consider the even-parity pairing Hamiltonian of Eq.~\eqref{eq:Delta_2band}. 
We further assume that only the $+$ band has a normal-state Fermi surface and the energy scales of the inter-band and intra-band pairings are much smaller than the energy splitting $|E_{\kk,+}-E_{\kk,-}|$ in the vicinity of the Fermi surface. 
Then, the BFSs appear when the intra-band pair potential, $\psi_{\kk,+}$, has nodes and the inter-band pair potentials, $\psi_{\kk,I}$ and $\bm d_\kk$, break TRS~\cite{Agterberg2017} (See Appendix~\ref{sec:app_Pf} for the details).
On the other hand, when only the $+$ band has a Fermi surface, we often approximate the system as a pseudo-spin-1/2 single-band system and consider only the intra-band pairing potential, $\psi_{\kk,+}$. 
The corresponding excitation spectrum $\sqrt{(E_{\kk,+}-\mu)^2+|\psi_{\kk,+}|^2}$ may have nodal points and lines on the Fermi surface along the direction of $\psi_{\kk,+}=0$ when $\psi_{\kk,+}$ has \textit{d}-wave symmetry, but there is no nodal surface.
The BFSs appear only when the inter-band coupling exists and breaks TRS: In such a case, the polarization~\eqref{eq:spin_polarization} works as an effective magnetic field for the $+$ band and leads to the splitting of the degenerate BdG spectrum~\cite{Brydon2018}. This is the origin of the BFSs. 
Hence, the existence of the TRS-breaking inter-band pairing is crucial for the appearance of BFSs, and it deforms the point and line nodes of $\psi_{\kk,+}$ to the BFSs.
In the following sections, we investigate how the properties of the nodal \textit{d}-wave pairing states in a single-band system change when the nodes deform into BFSs by taking into account the inter-band pairing.

\section{Spin-\texorpdfstring{$\bm{3/2}$}{} System with cubic symmetry}
\label{sec:luttinger_kohn}
As a concrete model for a two-band system with TRS, we consider spin-3/2 electrons in the Luttinger-Kohn model~\cite{Luttinger1955},
where four internal degrees of freedom reflect the total angular momentum $j=3/2$ combined from spin angular momentum $s=1/2$ and orbital angular momentum $l=1$ of the $p$-band.
Spin-3/2 electrons are known to appear, for example, in the $\Gamma_8$ band of cubic crystals with strong spin-orbit interactions.

In this section, we introduce an $O_h$-symmetric BdG Hamiltonian in the $j=3/2$ basis.
When we describe the internal degrees of freedom using the $j=3/2$ basis, 
we use quantities with bars, such as $\bar{\bm c}_\kk=(\bar{c}_{\kk,3/2},\bar{c}_{\kk,1/2},\bar{c}_{\kk,-1/2}, \bar{c}_{\kk,-3/2})^{\rm T}$, $\bar{H}_0(\kk)$, and $\bar{\Delta}(\kk)$,
which are related to the ones in the band basis via a unitary transformation.

\subsection{Normal-state Hamiltonian}
\begin{figure}[tb]
\begin{center}
    \input{dispersion}
    \caption{
    Schematic energy dispersion of $j=3/2$ electrons in the Luttinger-Kohn model around ${\bm k}={\bm 0}$ for $0<m_1=m_2<|m_0|$ and $\mu>0$. There are two doubly-degenerate bands, $\xi_{\bm k,+}$ and $\xi_{\bm k,-}$ [Eq.~\eqref{eq:xi_k_pm}], and only the $+$ bands have Fermi surfaces.
    }
    \label{fig:normal_band}
\end{center}
\end{figure}
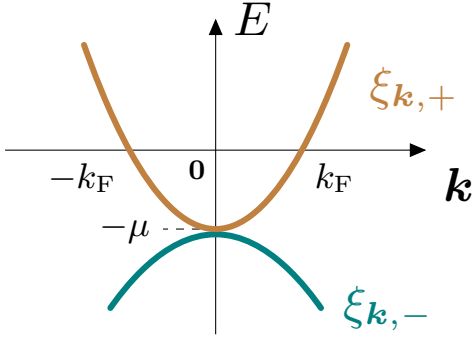

The normal part of an $O_h$-symmetric Hamiltonian in a continuum up to the second order of momentum is given by~\cite{Luttinger1955}
\begin{align}
    \bar{H}_0(\kk)=&\left(\frac{\hbar^2k^2}{2m_0}-\mu\right)\mathbbm{1}_4\nonumber\\
    &+\frac{3\hbar^2}{4m_1} \sum_{\nu=1,2,3}\mathscr{Y}_{\nu}(\kk)\mathscr{Y}_{\nu}(\JJ)\nonumber\\
    &+\frac{3\hbar^2}{4m_2} \sum_{\nu=4,5}\mathscr{Y}_{\nu}(\kk)\mathscr{Y}_{\nu}(\JJ),
\label{eq:H0_Oh}
\end{align}
where $\mathscr{Y}_\nu(\bm A)\ (\nu=1,2,3,4,5)$ are the rank-2 irreducible tensors constitute of a three-dimensional vector ${\bm A}=(A_x,A_y,A_z)$:
\begin{subequations}
\begin{align}
    \mathscr{Y}_{1}(\bm A)&=\frac{1}{\sqrt{3}}(A_xA_y+A_yA_x),    \label{eq:Y_1}\\
    \mathscr{Y}_{2}(\bm A)&=\frac{1}{\sqrt{3}}(A_yA_z+A_zA_y),\label{eq:Y_2}\\
    \mathscr{Y}_{3}(\bm A)&=\frac{1}{\sqrt{3}}(A_zA_x+A_xA_z),\label{eq:Y_3}\\
    \mathscr{Y}_{4}(\bm A)&=\frac{1}{\sqrt{3}}(A_x^2-A_y^2),\label{eq:Y_4}\\
    \mathscr{Y}_{5}(\bm A)&=\frac{1}{3}(2A_z^2-A_x^2-A_y^2),\label{eq:Y_5}
\end{align}
\label{eq:Y-tensors}
\end{subequations}
and $\JJ=(J_x,J_y,J_z)$ is the vector of the spin-3/2 matrices.
A set of $\gamma^\nu\equiv\mathscr{Y}_\nu(\JJ)$ corresponds to the anticommuting Dirac matrices~\cite{Brydon2018}. 
The specific forms of the matrices $J_{x,y,z}$ and $\gamma^\nu$ will be given in Appendix~\ref{sec:app_matrix}.
Since $\mathscr{Y}_{1,2,3}(\bm A)$ and $\mathscr{Y}_{4,5}(\bm A)$ are the basis of irreducible representations $T_{2g}$ and $E_g$, respectively, of the point group $O_h$, 
the effective masses $m_1$ and $m_2$ are independent constants.
For the sake of simplicity, we choose $m_1=m_2$ in the following calculations, with which the Hamiltonian~\eqref{eq:H0_Oh} becomes invariant under $O(3)$. Then, the eigenvalue of the Hamiltonian~\eqref{eq:H0_Oh} is given by
\begin{align}
    \xi_{\kk,\pm}&=\frac{\hbar^2k^2}{2M_\pm} -\mu, \label{eq:xi_k_pm}\\
    M_{\pm}&=\frac{1}{1/m_0\pm 1/m_1},
\end{align}
which gives spherically symmetric normal-state Fermi surfaces.
We further choose $0<m_1<|m_0|$ and $\mu>0$ so that only the $+$ band has a Fermi surface as shown in Fig.~\ref{fig:normal_band}.

\subsection{Pairing Hamiltonian}
\label{sec:PairingHamiltonian}
The pair potential for a spin-3/2 system is distinct from that for a simple two-band system in that the pairing states can be classified according to the total angular momentum of a Cooper pair $J_\textrm{pair}$.
For the case of even-parity pairing, the internal degrees of freedom of a Cooper pair can be
singlet ($J_\textrm{pair}=0$), quintet ($J_\textrm{pair}=2$), or a superposition of them~\cite{Brydon2016}.
By using the unitary part of the time-reversal operator
\begin{align}
    U_T
    =\exp\left(-i\pi J_y\right)
    =\begin{pmatrix}
        0 & 0 & 0 & -1\\
        0 & 0 & 1 & 0\\
        0 & -1 & 0 & 0\\
        1 & 0 & 0 & 0
        \end{pmatrix}    
\end{align}
and the rank-2 irreducible tensors $\mathscr{Y}_j(\JJ)$,
we write a general form of the pairing part of the Hamiltonian as
\begin{align}
    \bar{\Delta}(\kk)=\eta_0(\kk)U_T+\sum_{\nu=1}^5\eta_\nu(\kk)\mathscr{Y}_\nu(\JJ)U_T,
\label{eq:Delta3/2_expand}
\end{align}
where the first and second terms corresponds to the $J_\textrm{pair}=0$ and $J_\textrm{pair}=2$ pair potentials, respectively, and $\eta_0(\kk)$ and $\eta_\nu(\kk)$ are even functions of $\kk$.

Under the cubic symmetry, the pair potential is further divided into three irreducible representations $A_{1g}$, $T_{2g}$, and $E_g$, respectively given by
\begin{subequations}
\begin{align}
    A_{1g}:\ \  \bar{\Delta}(\kk)&=\eta_0(\kk)U_T,\label{eq:IR_states_spin3/2_a}\\
    T_{2g}:\ \  \bar{\Delta}(\kk)&=\sum_{\nu=1,2,3}\eta_\nu(\kk)\mathscr{Y}_\nu(\JJ)U_T,\label{eq:IR_states_spin3/2_b}\\
    E_g:\ \  \bar{\Delta}(\kk)&=\sum_{\nu=4,5}\eta_\nu(\kk)\mathscr{Y}_\nu(\JJ)U_T\label{eq:IR_states_spin3/2_c}.
\end{align}
\label{eq:IR_states_spin3/2}%
\end{subequations}
Because we are interested in TRS-broken states, we consider the following three cases of $J_\textrm{pair}=2$ pairs:
\begin{subequations}
\begin{align}
    T_{2g} (0,i,1):&\ \bm\eta=\Delta_0(0,0,i,1,0,0),
    \label{eq:3states_spin3/2_a}\\
    T_{2g} (1,\omega,\omega^2):&\ \bm\eta=\Delta_0(0,1,\omega,\omega^2,0,0),
    \label{eq:3states_spin3/2_b}\\
    E_g (1,i):&\ \bm\eta=\Delta_0(0,0,0,0,1,i),
    \label{eq:3states_spin3/2_c}
\end{align}
\label{eq:3states_spin3/2}%
\end{subequations}
where $\bm\eta(\kk)\equiv(\eta_0(\kk),\eta_1(\kk),\cdots,\eta_5(\kk))$, $\omega\equiv e^{i2\pi/3}$,
and we consider only the simplest cases of $\kk$-independent gap functions.
We note that these states spontaneously break the cubic symmetry in addition to the TRS but are known to be stable solutions of the fourth-order Ginzburg–Landau theory~\cite{SigristUeda1991}

\subsection{Description in the band basis}
\label{sec:Description_in_Band}
We introduce a unitary transformation $U(\bm k)$ that diagonalizes the normal Hamiltonian
\begin{align}
    U^\dagger(\kk) \bar{H}_0(\kk)U(\kk)=\Diag[\xi_{\kk +},\xi_{\kk +},\xi_{\kk -},\xi_{\kk -}],
\end{align}
and describe the system in the band basis.
Because the band basis $\bm c_\kk$ is related to the $j=3/2$ bases $\bar{\bm c}_\kk$ as $\bar{\bm c}_\kk=U(\kk)\bm c_\kk$ and $\bar{\bm c}^\dagger_\kk=U^*(\kk)\bm c_\kk^\dagger$, the pairing part of the Hamiltonian is written in the band basis as
\begin{align}
    \Delta(\kk)=U^\dagger(\kk)\bar{\Delta}(\kk)U^*(\kk).
    \label{eq:Ud_Delta_Uc}
\end{align}
According to the discussion in Sec.~\ref{sec:BFS}, this $\Delta(\kk)$ is written in the form of Eq.~\eqref{eq:Delta_2band}.

We note here that because the unitary transformation $U(\kk)$ depends on $\kk$, $\Delta(\kk)$ in the band basis becomes $\kk$-dependent even when $\bar{\Delta}(\kk)$ is $\kk$-independent. In particular, in the present cases of $J_\textrm{pair}=2$ under the $O(3)$ symmetric normal-state Hamiltonian, the effective intra-band pair potential $\psi_{\kk,+}$ has \textit{d}-wave symmetry.
This can be proved as follows. First, $U(\kk)$ for the present system is given by
\begin{align}
    U(\kk)=e^{i\pi/2}e^{-i J_z \phi}e^{-i J_y\theta}
    \begin{pmatrix} 1 & 0 & 0 & 0 \\ 0 & 0 & 0 & 1 \\ 0 & 0 & 1 & 0 \\ 0 & 1 & 0 & 0\end{pmatrix}
\end{align}
where $\theta=\arccos{k_z/k}$ and $\phi=\arg(k_x+ik_y)$.
This unitary operator transforms $U_T$ and $\mathscr{Y}_\nu(\JJ)U_T$ in the pair potential as
\begin{subequations}
\begin{align}
    \mathcal{P}_+[U^\dagger(\kk)U_TU^*(\kk)]&=i\sigma_y\\
    \mathcal{P}_+[U^\dagger(\kk)\mathscr{Y}_\nu(\JJ)U_TU^*(\kk)]&=\frac{3}{2}\mathscr{Y}_\nu(\kk)i\sigma_y
    \label{eq:pair_transform}
\end{align}
\label{eq:pair_transform_all}
\end{subequations}
where $\mathcal{P}_+$ means the projection onto the $+$ band, i.e., taking the upper left $2\times 2$ matrix of the $4\times 4$ matrix.
Equation~\eqref{eq:pair_transform} clearly shows that the anisotropic pairing in the spin space of the $J_\textrm{pair}=2$ system
is transcribed to the anisotropic $\kk$-dependence of the intra-band pairing. 
From Eqs.~\eqref{eq:Delta3/2_expand}, \eqref{eq:Ud_Delta_Uc}, and \eqref{eq:pair_transform_all}, we obtain the intra-band pair potential as
\begin{align}
\mathcal{P}_+[\Delta(\bm k)]&=\psi_{\bm k,+}i\sigma_y
\label{eq:projection_to_spin1/2}
\end{align}
with
\begin{align}
\psi_{\bm k,+}&=\eta_0(\bm k)+\frac{3}{2}\sum_{\nu=1}^5 \eta_\nu(\bm k)\mathscr{Y}_{\nu}(\bm k).
\label{eq:intra-band_pair}
\end{align}
Accordingly, the spin-singlet pair potentials corresponding to the states in Eq.~\eqref{eq:3states_spin3/2} are respectively given by
\begin{subequations}
\begin{align}
    T_{2g} (0,i,1):&\ \psi_{\kk,+}=\Delta_0 \sqrt{3} k_z(k_x+ik_y),
    \label{eq:3states_spin1/2_a}\\
    T_{2g} (1,\omega,\omega^2):&\ \psi_{\kk,+}=\Delta_0 \sqrt{3}(k_x k_y+\omega k_y k_z + \omega^2 k_z k_x),
    \label{eq:3states_spin1/2_b}\\
    E_g (1,i):&\ \psi_{\kk,+}=\Delta_0e^{-i\pi/6}(k_x^2+\omega^2 k_y^2 + \omega k_z^2).
    \label{eq:3states_spin1/2_c}
\end{align}
\label{eq:3states_spin1/2}%
\end{subequations}
Here, the intra-band pair potential for $T_{2g} (0,i,1)$ is the 3D chiral \textit{d}-wave state, which is a possible candidate for the superconducting state in \ce{Sr2RuO4}~\cite{Agterberg2020},
whereas that for $T_{2g} (1,\omega,\omega^2)$ is the 3D cyclic \textit{d}-wave state, which is predicted to appear in neutron stars~\cite{MizushimaNitta}.
Since BFS is generated by inter-band pseudospin-triplet pair (discussed in Appendix~\ref{sec:app_Pf}) , the nodal structure is simply generated by Eq.~\eqref{eq:3states_spin1/2} which is only intra-band pair potential.

All of the three pairing states in Eq.~\eqref{eq:3states_spin1/2} have nodal gap structures, indicating the appearance of the BFSs in the original spin-3/2 system.
In addition, because the \textit{d}-wave pairing states in Eq.~\eqref{eq:3states_spin1/2} are topological, zero-energy surface flat bands and surface arc states appear in a particular direction of the interface, which resonantly increases the tunnel conductance at zero bias~\cite{Kobayashi2015,Tamura2017}.
We, therefore, focus on how such a behavior of the tunnel conductance around zero bias changes when the line nodes and point nodes deform to the BFSs
by comparing the $J_{\rm pair}=2$ pairing states in Eq.~\eqref{eq:3states_spin3/2} and
the spin-singlet \textit{d}-wave pairing states in Eq.~\eqref{eq:3states_spin1/2}.

\section{Setup and methods}
\label{sec:setup}

\begin{figure}[htbp]
  \begin{center}
    \includegraphics[width=\linewidth]{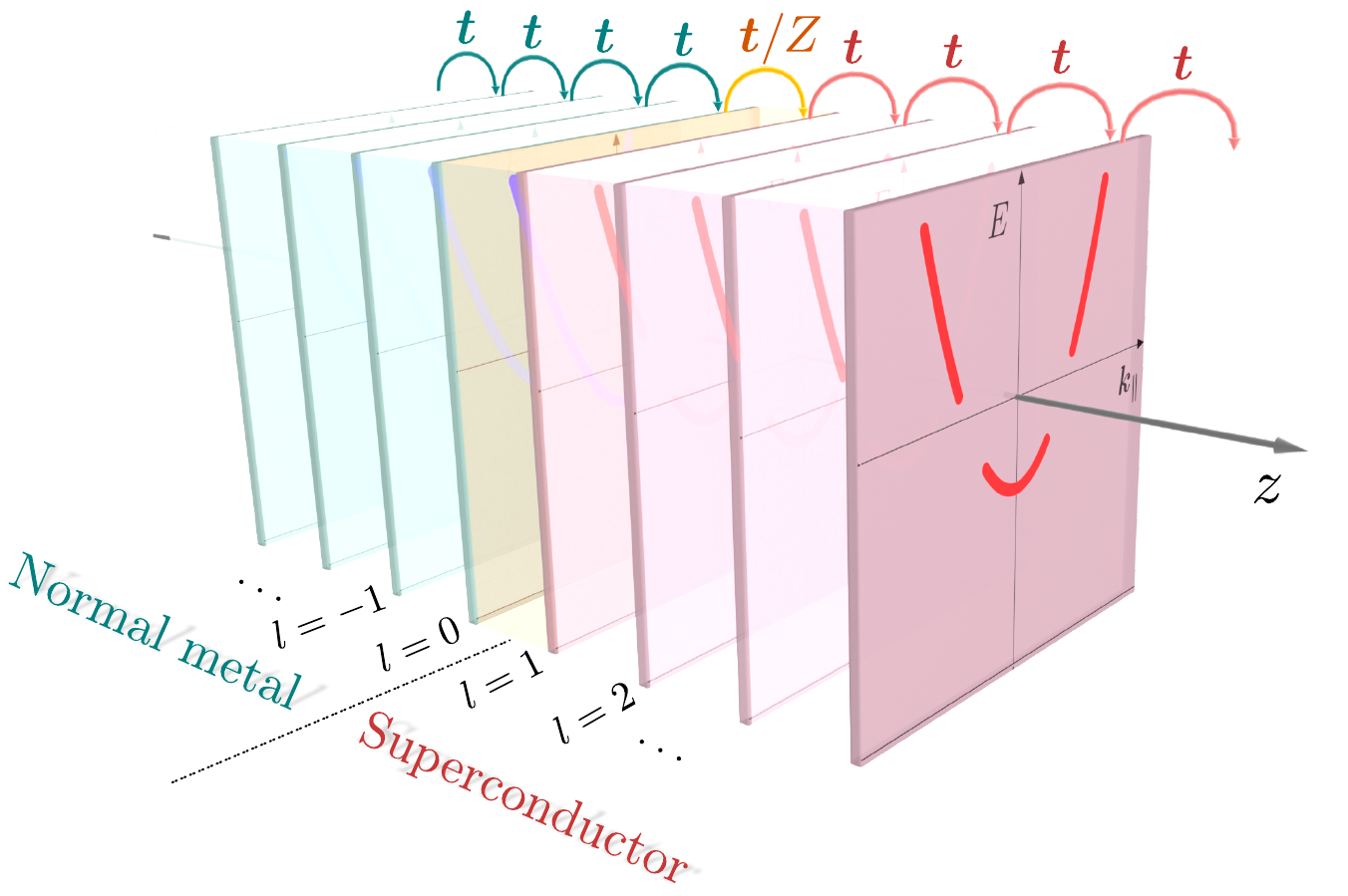}
    \caption{Schematic illustration of a normal metal/superconductor (N/S) junction. We use the same normal state Hamiltonian for both sides and introduce superconducting pair potential in the $z>0$ side. 
    The transfer matrix at the interface is multiplied by $1/Z$ with $Z$ being the barrier parameter defined in Eq.~\eqref{eq:interface_hamiltonian}.
    }
    \label{fig:setup}
  \end{center}
\end{figure}

We consider a normal metal/superconductor (N/S) junction as schematically shown in Fig.~\ref{fig:setup}
and compare the results for $j=3/2$ and $j=1/2$ systems having the same symmetry.
For the case of the $j=3/2$ system, we use the same normal-state Hamiltonian $\bar{H}_0(\kk)$ given by Eq.~\eqref{eq:H0_Oh} both in the normal metal and the superconductor.
The pair potential is zero in the normal metal, whereas that in the superconductor is one of Eq.~\eqref{eq:3states_spin3/2}.
For the case of the $j=1/2$ system, we use
the normal state Hamiltonian $\xi_{\kk,+}\otimes \mathbbm{1}_2$, where $\xi_{\kk,+}$ is given by Eq.~\eqref{eq:xi_k_pm} with $M_+$ and $\mu$ being the same as those in the $j=3/2$ model.
The pair potential is zero in the normal metal, whereas that in the superconductor has \textit{d}-wave symmetry given by one of Eq.~\eqref{eq:3states_spin1/2}.

To calculate the Green's function at the interface, we use the recursive Green's function method~\cite{Umerskii, TamuraTakagi}, which is applicable for a tight-binding model along the perpendicular direction to the interface.
We choose the $z$ axis perpendicular to the interface and replace the terms including $k_z$ in the BdG Hamiltonian as
\begin{subequations}
\begin{align}
    k_{x,y}k_z&\to \frac{1}{a}k_{x,y}\sin k_za, \\
    k_z^2&\to \frac{2}{a^2}(1-\cos k_za),
\end{align}
\label{eq:kz_trans}%
\end{subequations}
with $a$ being the lattice spacing along the $z$ axis,
so that the Hamiltonian is rewritten 
in terms of the Fourier-transformed operators with respect to $k_z$,  ${\bm f}_{l,\kk_\perp}=\sum_{k_z}e^{ik_z al}\bar{\bm c}_{\kk}$ with $\kk_\perp=(k_x,k_y)$. 
The resulting Hamiltonian with an interface between $l=0$ and $1$ is given by
$\mathcal{H}=\sum_{\kk_\perp}\mathcal{H}(\kk_\perp)$ with
\begin{align}
    \mathcal{H}(\kk_\perp)=&\sum_{l\le -1}h^{\rm N}_{\rm L}(l,\kk_\perp)     
    +h_{\rm I}(\kk_\perp) \nonumber\\
    &+\sum_{l\ge 1}\left[h^{\rm N}_{\rm R}(l,\kk_\perp)+h^{\rm S}_{\rm R}(l,\kk_\perp)\right],
\end{align}
where $h^{\rm N}_{\rm L(R)}(l,\kk_\perp)$, $h^{\rm S}_{\rm R}(l,\kk_\perp)$, $h_{\rm I}(\kk_\perp)$ are the normal-state Hamiltonian in the left-hand (right-hand) side of the interface, the pairing Hamiltonian in the right-hand side, and the Hamiltonian at the interface, respectively.
We use the same normal-state Hamiltonian:
\begin{align}
h_{\rm L}^{\rm N}(l,\kk_\perp)=&h_{\rm R}^{\rm N}(l,\kk_\perp)\nonumber\\
=&-(\bm f_{l+1,\kk_\perp}^\dagger)^{\rm T} t(\kk_\perp) \bm f_{l,\kk_\perp}\nonumber\\
&-(\bm f_{l,\kk_\perp}^\dagger)^{\rm T} t^\dagger(\kk_\perp) \bm f_{l+1,\kk_\perp}\nonumber\\
&+(\bm f_{l,\kk_\perp}^\dagger)^{\rm T} u(\kk_\perp)\bm f_{l,\kk_\perp},
\label{eq:normal-Hamiltonian}
\end{align}
where $t(\kk_\perp)$ and $u(\kk_\perp)$ are the $\kk_\perp$-dependent hopping matrix and the on-site potential matrix, respectively, derived from the continuum model with the replacement in Eq.~\eqref{eq:kz_trans}.
Here, $t(\kk_\perp)$ and $u(\kk_\perp)$ are $4\times 4$ ($2\times 2$) matrices for the $j=3/2$ ($j=1/2$) system,
whose detailed forms are given in Appendix~\ref{sec:app_tb}.
We introduce barrier parameter $Z$ and multiply $1/Z$ to the transfer matrix at the interface, i.e., we use the following interface Hamiltonian:
\begin{align}
    h_{\rm I}(\kk_\perp)=&-\frac{1}{Z}\left[(\bm f_{1,\kk_\perp}^\dagger)^{\rm T} t(\kk_\perp) \bm f_{0,\kk_\perp}+(\bm f_{0,\kk_\perp}^\dagger)^{\rm T} t^\dagger(\kk_\perp) \bm f_{1,\kk_\perp}\right]\nonumber\\
    &+(\bm f_{0,\kk_\perp}^\dagger)^{\rm T} u(\kk_\perp) \bm f_{0,\kk_\perp}.
    \label{eq:interface_hamiltonian}
\end{align}

The pairing Hamiltonian consists of the on-site and off-site pairing terms:
\begin{align}
    h_{\rm R}^{\rm S}(l,\kk_\perp)=&(\bm f_{l,\bm k_\perp}^\dagger)^{\rm T} \bar{\Delta}_{\rm on}(\kk_\perp) \bm f_{l,-\bm k_\perp}^\dagger\nonumber\\
    &+ (\bm f_{l,\bm k_\perp}^\dagger)^{\rm T} \bar{\Delta}_{\rm off}(\kk_\perp) \bm f_{l+1,-\bm k_\perp}^\dagger\nonumber\\
    &+ \textrm{H. c.},
\end{align}
where H. c. stands for the Hermitian conjugate.
In the case of the $j=3/2$ system with the $\bm k$-independent pair potentials in Eq.~\eqref{eq:3states_spin3/2}, only the on-site pair potential exists, i.e., 
\begin{align}
\bar{\Delta}_{\rm on}(\kk_\perp)=\bar{\Delta},\ \ \bar{\Delta}_{\rm off}=0,
\end{align}
where $\bar{\Delta}$ is given by Eq.~\eqref{eq:Delta3/2_expand} with $\kk$-independent $\eta_j$'s in Eq.~\eqref{eq:3states_spin3/2}. On the other hand, in the $j=1/2$ system, the pair potential has \textit{d}-wave symmetry, and $h_{\rm R}^{\rm S}$ includes both the on-site and off-site pairing terms.
The concrete forms of $\bar{\Delta}_{\rm on}(\kk_\perp)$ and $\bar{\Delta}_{\rm off}(\kk_\perp)$ corresponding to the pair potentials in Eq.~\eqref{eq:3states_spin1/2} is given in Appendix~\ref{sec:app_tb}.
 
The retarded (advanced) Green's function for the BdG Hamiltonian is written as
\begin{align}
    \mathcal{G}^{\rm R(A)}_{ll'}(\epsilon,\kk_\perp)&=\left[\epsilon+(-)i\delta_\epsilon-\mathcal{H}(\kk_\perp)\right]^{-1}_{ll'}\\
    &=\begin{bmatrix}
    G^{\rm R(A)}_{ll'}(\epsilon,\kk_\perp) & F^{\rm R(A)}_{ll'}(\epsilon,\kk_\perp) \\[1mm] \tilde{F}^{\rm R(A)}_{ll'}(\epsilon,\kk_\perp) & \tilde{G}^{\rm R(A)}_{ll'}(\epsilon,\kk_\perp)
    \end{bmatrix}.
\end{align}
Here, $G^{\rm R(A)}_{ll'}, \tilde{G}^{\rm R(A)}_{ll'}, F^{\rm R(A)}_{ll'}$ and $\tilde{F}^{\rm R(A)}_{ll'}$ are $4\times 4$ ($2\times 2$) matrices for the $j=3/2$ ($j=1/2$) system.
We calculate the Green's function at the interface by using the recursive Green's function method.

To see the effects of the BFSs and topological surface states on the charge conductance,
we first calculate the SDOSs of the $j=1/2$ and $j=3/2$ superconductors and compare them with the LDOSs deep inside the superconductors.
Here, SDOS, $\rho_{\rm S}(\epsilon)$, and momentum-resolved SDOS, $\rho_{\rm S}(\epsilon,\kk_\perp)$, are calculated via
    \begin{subequations}
    \begin{align}
    \rho_{\rm S}(\epsilon)&=\sum_{\kk_\perp}\rho_{\rm S}(\epsilon,\kk_\perp),\\
    \rho_{\rm S}(\epsilon,\kk_\perp)&=-\frac{1}{\pi}{\rm Im}[G^{\rm R}_{11}(\epsilon,\kk_\perp)],
    \end{align}
    \label{eq:SDOS}%
    \end{subequations}
under the Hamiltonian with $h_{\rm L}^{\rm N}(l,\kk_\perp)=h_{\rm I}(\kk_\perp)=0$.
Similarly, bulk LDOS, $\rho_\textrm{S}^\textrm{(bulk)}(\epsilon)$, inside the superconductor is given by the same form as Eq.~\eqref{eq:SDOS} but 
with $G_{11}^{\rm R}(\epsilon,\kk_\perp)$ calculated by adding the same pairing Hamiltonian in both sides and choosing $Z=1$.
We also define the normal-state version of these quantities, $\rho_\textrm{N}(\epsilon)$ and $\rho_\textrm{N}^\textrm{(bulk)}(\epsilon)$, which are calculated using the same code but with choosing $\Delta_0=0$.

Next, we calculate the charge conductance by utilizing the Lee-Fisher formula~\cite{Lee1981}.
Here, we use the generalized one for the case with internal degrees of freedom~\cite{Inoue_text}.
We describe the formula in the Nambu space (see Appendix~\ref{sec:app_LF} for details) and obtain 
\begin{align}
\Gamma(\epsilon)=\frac{e^2}{2h}&\Tr\sum_{\kk_\perp}\nonumber\\
\bigg[
&-\bar{\mathcal{G}}_{01}(\epsilon,\kk_\perp){T}(\kk_\perp)\bar{\mathcal{G}}_{01}(\epsilon,\kk_\perp){T}(\kk_\perp)\nonumber\\
&-\bar{\mathcal{G}}_{10}(\epsilon,\kk_\perp){T}^\dagger(\kk_\perp)\bar{\mathcal{G}}_{10}(\epsilon,\kk_\perp){T}^\dagger(\kk_\perp)\nonumber\\
&+\bar{\mathcal{G}}_{11}(\epsilon,\kk_\perp){T}(\kk_\perp)\bar{\mathcal{G}}_{00}(\epsilon,\kk_\perp){T}^\dagger(\kk_\perp)\nonumber\\
&+\bar{\mathcal{G}}_{00}(\epsilon,\kk_\perp){T}^\dagger(\kk_\perp)\bar{\mathcal{G}}_{11}(\epsilon,\kk_\perp){T}(\kk_\perp)\bigg],
\label{eq:Lee-Fisher}
\end{align}
where  
\begin{align}
    \bar{\mathcal{G}}_{ll'}(\epsilon,\kk_\perp) &= \frac{1}{2i}[\mathcal{G}^{\rm A}_{ll'}(\epsilon,\kk_\perp)-\mathcal{G}^{\rm R}_{ll'}(\epsilon,\kk_\perp)],\\
    T(\kk_\perp)&=\begin{bmatrix} \bm t(\kk_\perp) & \bf 0 \\ \bf 0 & \bm t^*(-\kk_\perp) \end{bmatrix}.
    \label{eq:Tkk}
\end{align}
To see the effect of the BFSs on the tunneling spectroscopy, we calculate the conductance $\Gamma_{\rm S}(\epsilon)$ for the N/S junction 
and normalize it with that for the normal metal/normal metal junction, $\Gamma_{\rm N}(\epsilon)$, which is calculated by choosing $\Delta_0=0$.

In the following numerical calculations, we use $\mu$ as the unit of energy and choose
\begin{subequations}
\begin{align}
    \frac{\hbar^2}{m_0a^2}&=\frac{\mu}{6},\\
    \frac{\hbar^2}{m_1a^2}&=\frac{\hbar^2}{m_2a^2}=4\mu.
\end{align}
\label{eq:TB_parameters}%
\end{subequations}
The corresponding Fermi wave number for the $+$ band in the continuum model is 
$k_{\rm F}a=\sqrt{2M_+\mu}/\hbar=2\sqrt{3}/5\sim 0.69$.
We choose the infinitesimal value $\delta_\epsilon$ as $\delta_\epsilon/\mu = 5.0\times 10^{-5}$ for LDOS and SDOS calculations and $\delta_\epsilon/\mu = 1.0\times 10^{-8}$ for conductance calculations.

\section{Results}
\label{sec:results}
In this section, we show the results of the SDOS and the charge conductance for each pairing state of Eqs.~\eqref{eq:3states_spin3/2} and \eqref{eq:3states_spin1/2}, which are briefly summarized as follows. The \textit{d}-wave pairing states in Eqs.~\eqref{eq:3states_spin1/2_a}, \eqref{eq:3states_spin1/2_b}, and \eqref{eq:3states_spin1/2_c} have topological zero-energy flat-band SABSs, topological SABSs with surface arc states, and no SABS, respectively, on the (001) surface. 
When the point nodes and line nodes deform to BFSs in the $J_\textrm{pair}=2$ pairing states, the topological SABSs shift from zero energy or move in the surface Brillouin Zone.
As a result, the ZEP of the SDOS and the ZBCP arising in the \textit{d}-wave states are blunted or split in the corresponding $J_\textrm{pair}=2$ states [$T_{2g} (0,i,1)$ and $T_{2g} (1,\omega,\omega^2)$ states in Eqs.~\eqref{eq:3states_spin3/2_a} and \eqref{eq:3states_spin3/2_b}, respectively].
On the other hand, the states on the BFSs have a nonzero contribution to the charge conductance at zero energy, as in the case of normal state conduction.
This contribution is visible in the absence of SABSs [$E_g(1,i)$ state in Eq.~\eqref{eq:3states_spin3/2_c}].

In the following sections, 
the results for \textit{d}-wave pairing states are independent of $\Delta_0/\mu$ when we scale the energy by $\Delta_0$.
We therefore show the numerical results only for $\Delta_0/\mu=0.2$ unless otherwise noted.
On the other hand, the results for the $J_\textrm{pair}=2$ pairing states in spin-3/2 systems exhibit a clear dependence on $\Delta_0/\mu$, which will be discussed in detail.

\subsection{\texorpdfstring{$\bm{T_{2g}\;(0,i,1)}$}{}}
\label{sec:t2g}
\textit{Short summary for $T_{2g}(0,i,1)$ state}:
When the line node in the 3D chiral \textit{d}-wave state changes into the BFS, the topologically protected zero-energy surface flat bands split and become dispersive. 
This change suppresses the sharp ZEP in the SDOS and the ZBCP, and the peak width becomes roughly in proportion to the square of the area of the BFS in the momentum space. 
In addition to the conventional Andreev reflection, the zero-energy states on the BFSs contribute to the charge conductance at $\epsilon=0$, leading to the unusual transmittance dependence of the normalized conductance.

\begin{figure}[htbp]
    \centering
    \includegraphics[width=\linewidth]{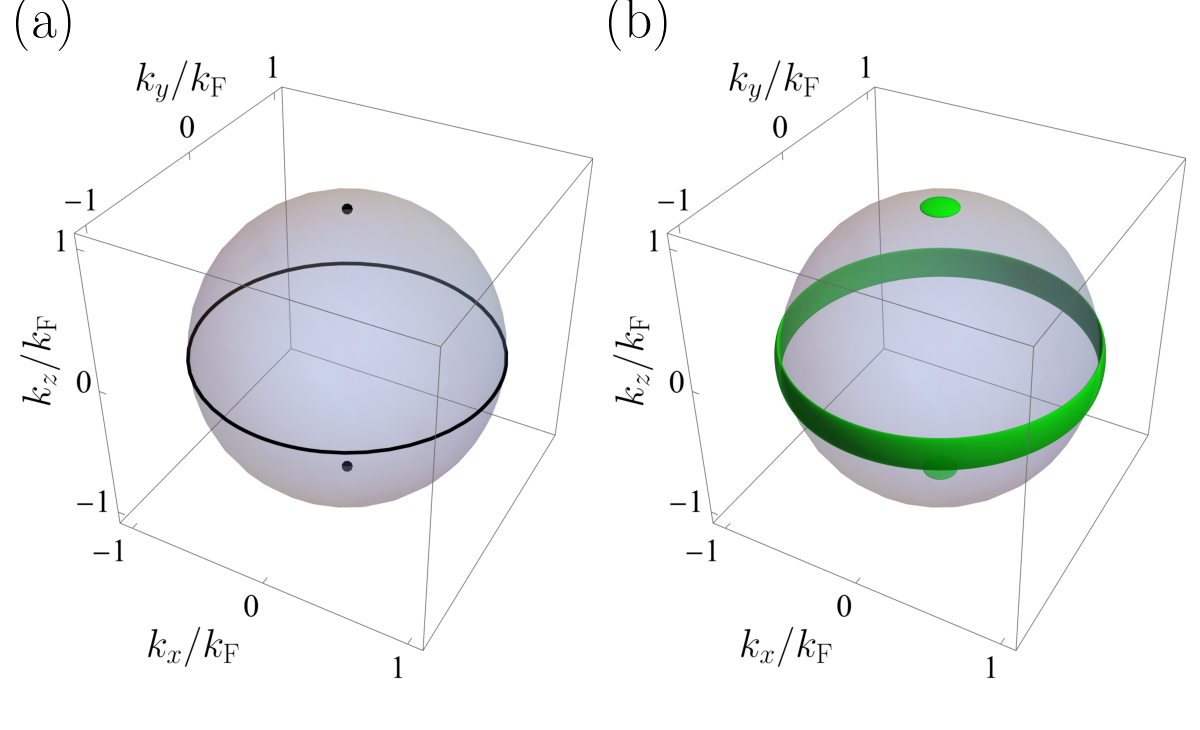}
    \caption{(a) Point and line nodes in 3D chiral \textit{d}-wave superconducting state [Eq.~\eqref{eq:3states_spin1/2_a}] and (b) Bogoliubov Fermi surfaces (BFSs) in spin-3/2 $T_{2g}(0,i,1)$ state  [Eq.~\eqref{eq:3states_spin3/2_a}] with $\Delta_{0}/\mu=0.2$. 
    }
    \label{fig:BFS_T2g@1i0}
\end{figure}

The pair potentials for the $T_{2g}(0,i,1)$ states of $j=1/2$ and $j=3/2$ systems are respectively given by Eqs.~\eqref{eq:3states_spin3/2_a} and \eqref{eq:3states_spin1/2_a}.
Here, the $j=1/2$ superconducting state is the 3D chiral \textit{d}-wave state which features one line node on the $k_xk_y$ plane and two point nodes on the $k_z$ axis as illustrated in Fig.~\ref{fig:BFS_T2g@1i0}(a).
The nodal regions in the $j=3/2$ system are extended to the BFSs as shown in Fig.~\ref{fig:BFS_T2g@1i0}(b), which is calculated by using Eq.~\eqref{eq:Pf_gen}.
The BFSs extend alongside the Fermi surface~\cite{Agterberg2017}: the line node expands into a circular ribbon shape, while the two point nodes stretch into a disc-like geometry.

\begin{figure}[htbp]
  \centering
  \includegraphics[width=\linewidth]{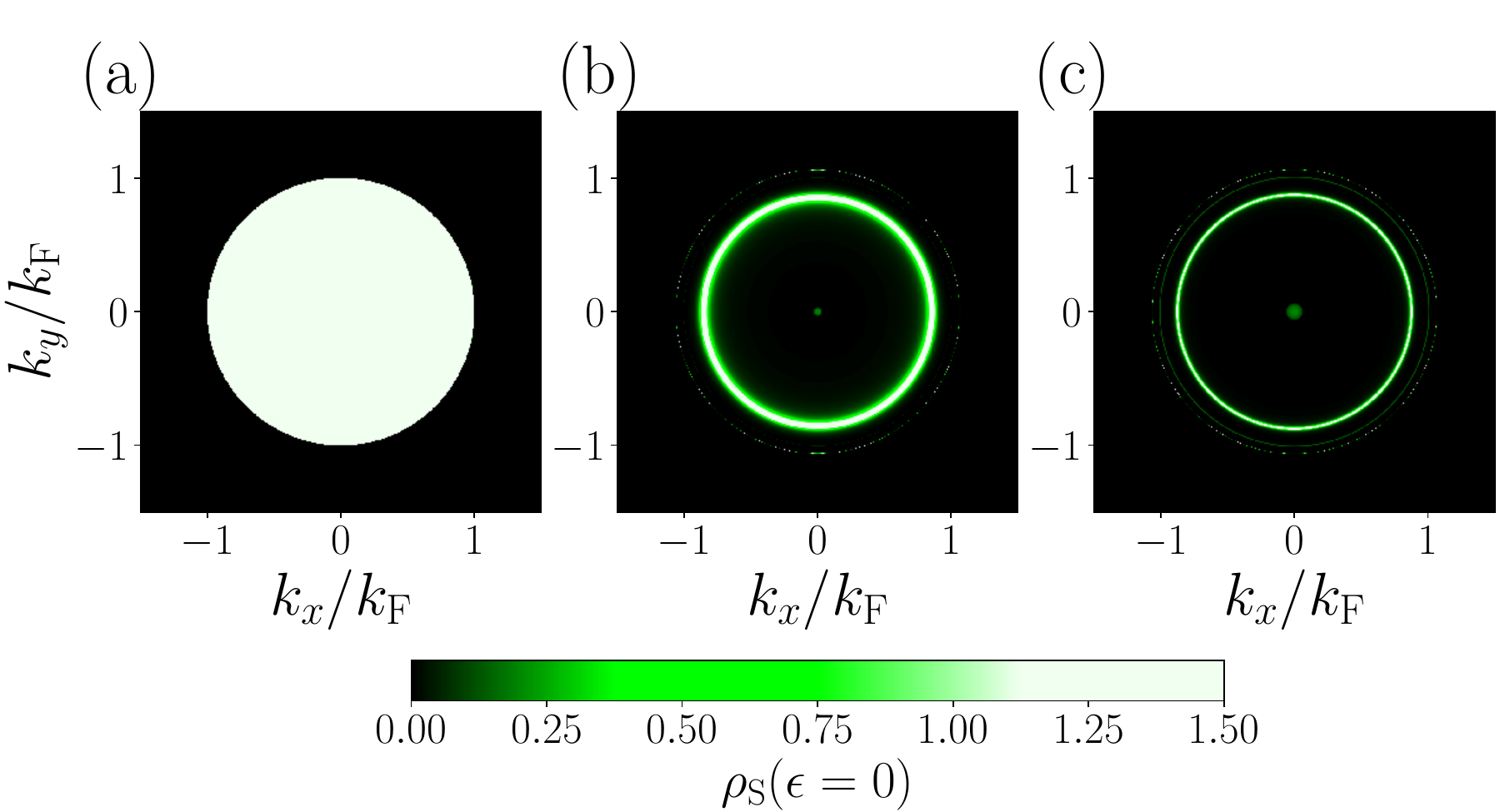}
    \caption{
    Momentum-resolved zero-energy SDOS for 3D chiral \textit{d}-wave state (a) and spin-3/2 $T_{2g} (0,i,1)$ state with $\Delta_0/\mu=0.05$ (b) and $0.1$ (c).
    Panel (a) does not depend on the value of $\Delta_0/\mu.$
    In panels (b) and (c), the zero-energy states remain on a ring, protected by the $CC_{2z}$ symmetry (see text).
    }
    \label{fig:SDOS_T2g@1i0_zero_ene}
\end{figure}
\begin{figure}[htbp]
  \begin{center}
    \includegraphics[width=\linewidth]{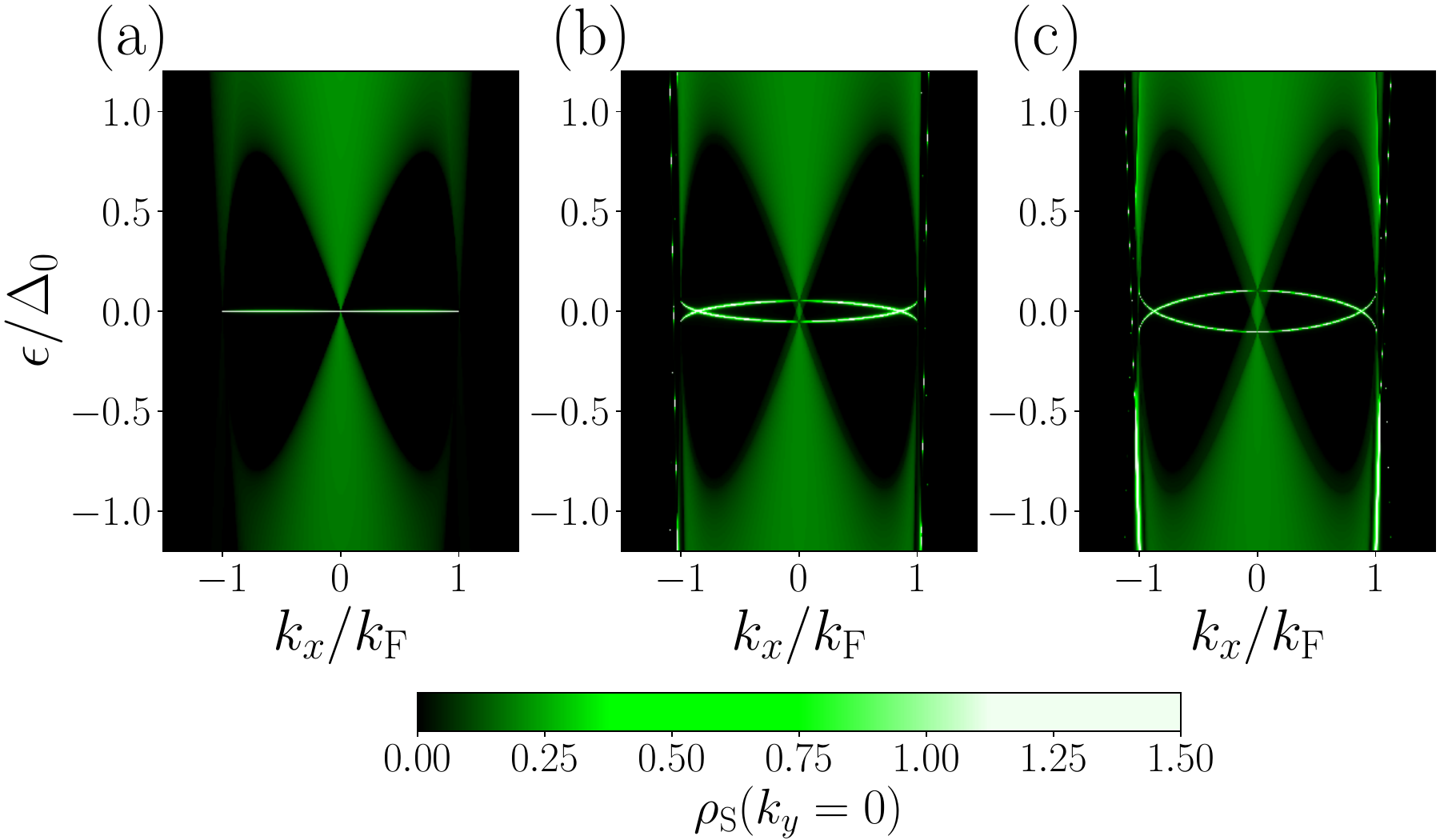}
  \caption{
    Momentum-resolved SDOS at $k_y=0$ for 3D chiral \textit{d}-wave state (a) and spin-3/2 $T_{2g} (0,i,1)$ state with $\Delta_0/\mu=0.05$ (b) and $0.1$ (c). 
    Panel (a) does not depend on the value of $\Delta_0/\mu.$
    In this and the following figures, the results for the \textit{d}-wave case are calculated for $\Delta_0/\mu=0.2$ unless otherwise noted. 
    The doubly-degenerate zero-energy flat bands in (a) split in the presence of BFSs in (b) and (c),
    where the energy splitting increases in proportion to $(\Delta_0/\mu)^2$.
    }
    \label{fig:SDOS_T2g@1i0}
    \end{center}
\end{figure}
Figures~\ref{fig:SDOS_T2g@1i0_zero_ene} and \ref{fig:SDOS_T2g@1i0} show the numerically obtained momentum-resolved SDOS at $\epsilon=0$ and $k_y=0$, respectively.
In the case of the 3D chiral \textit{d}-wave state [Figs.~\ref{fig:SDOS_T2g@1i0_zero_ene}(a) and \ref{fig:SDOS_T2g@1i0}(a)], doubly-degenerate zero-energy flat bands emerge, which are protected by the 1D winding number~\cite{Kobayashi2015}. 
On the other hand, in the case of the spin-3/2 $T_{2g} (0,i,1)$ state, the TRS-breaking pair potential lifts the degeneracy, and the surface states become dispersive as shown in Figs.~\ref{fig:SDOS_T2g@1i0_zero_ene}(b), \ref{fig:SDOS_T2g@1i0_zero_ene}(c), \ref{fig:SDOS_T2g@1i0}(b), and \ref{fig:SDOS_T2g@1i0}(c).
Because the spin polarization Eq.~\eqref{eq:spin_polarization}, which is proportional to $\Delta_0^2$, works as a pseudo-magnetic field, the maximum energy splitting at $k_x=k_y=0$ increases in proportion to $(\Delta_0/\mu)^2$~\cite{Brydon2018}.
Here, we note that the zero-energy states always remain on a ring on the surface Brillouin zone as shown in Figs.~\ref{fig:SDOS_T2g@1i0_zero_ene}(b) and \ref{fig:SDOS_T2g@1i0_zero_ene}(c) due to the $\kk$-dependence of the spin polarization. 
This is a distinctive difference from the case when we apply a uniform magnetic field to the 3D chiral \textit{d}-wave state, where the flat bands remain flat and shift in opposite directions. 
The existence of the ring-shaped zero-energy states is topologically protected by the Pfaffian associated with the $CC_{2z}$ symmetry of the BdG Hamiltonian with a boundary~\cite{Fang17, Ahn21}, where $C$ and $C_{2z}$ denote the particle-hole operator and the rotation by $\pi$ about the $z$ axis (perpendicular to the boundary), respectively: The Pfaffian of the BdG Hamiltonian in the presence of boundary, $H_{\kk_\perp}$, is well defined due to the relations
$(CC_{2z})^2=1$ and $(CC_{2z})H_{\kk_\perp}(CC_{2z})^{-1}=-H_{\kk_\perp}$.

\begin{figure}[htbp]
\centering
    \includegraphics[width=\linewidth]{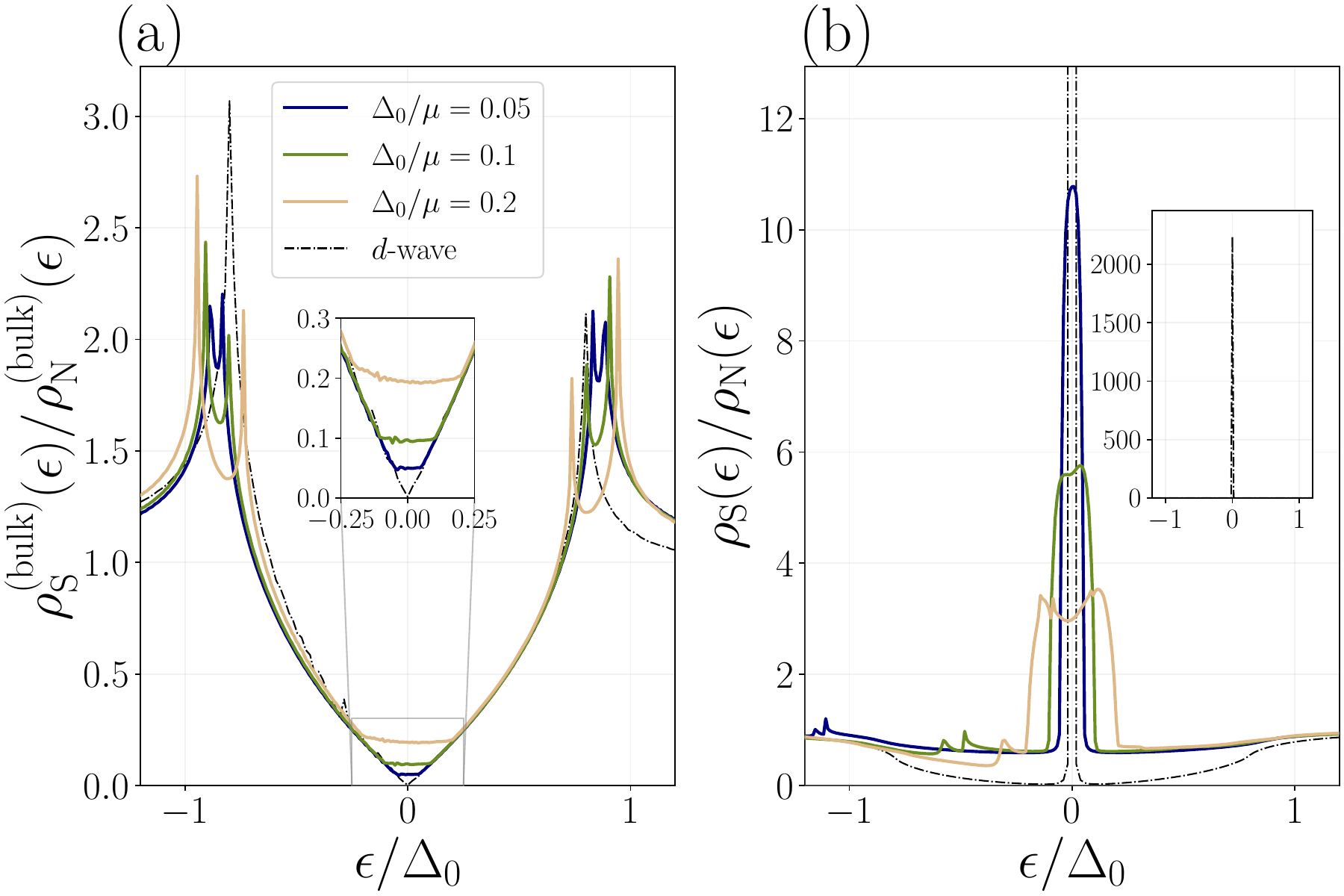}
    \caption{
    Bulk LDOS (a) and SDOS (b) of $T_{2g}(0,i,1)$ states normalized by their values at $\Delta_0=0$. 
    Inset in (a) is a magnified view around $\epsilon=0$, and inset in (b) is the SDOS of 3D chiral \textit{d}-wave state with an expanded vertical axis.
    The results for the \textit{d}-wave state are independent of the value of $\Delta_0/\mu.$
    (a) The BFSs contribute to the nonzero bulk LDOS at $\epsilon=0$.
    (b) The appearance of the BFSs blunts the sharp ZEP in the \textit{d}-wave case.
    }
    \label{fig:LDOS_T2g@1i0}
\end{figure}
To clarify the contributions of BFSs, we compare the SDOS with the bulk LDOS in Fig.~\ref{fig:LDOS_T2g@1i0}.
As known in the previous studies~\cite{Kobayashi2015,Tamura2017,YadaAndo2023,Suzuki2020},
the bulk LDOS in the 3D chiral \textit{d}-wave superconductor exhibits a V-shaped structure in the vicinity of $\epsilon=0$ because of the line node.
By contrast, the bulk LDOS in the spin-3/2 $T_{2g}(0,i,1)$ superconductor takes a constant value around $\epsilon=0$ proportional to $\Delta_0/\mu$.
This increase in the LDOS at $\epsilon=0$ is indeed attributed to the zero-energy states on the BFSs, whose area in the momentum space is proportional to $\Delta_0/\mu$~\cite{Agterberg2017}.
Furthermore, the energy range for the constant LDOS is proportional to $(\Delta_0/\mu)^{2}$, being consistent with the shift in the dispersion due to the spin-polarization.
The splitting of the coherence peak at $|\epsilon/\Delta_0|\sim 0.7$ is also explained by the dispersion shift due to the pseudo-magnetic field, and the width of the splitting is proportional to $(\Delta_0/\mu)^{2}$. 
[See Eq.~\eqref{eq:spin_polarization} and the last paragraph of Sec.~\ref{sec:BFS}]
When we terminate the system at the (001) surface, a sharp ZEP emerges in the SDOS of the 3D chiral \textit{d}-wave state due to the zero-energy surface flat bands  [Fig.~\ref{fig:LDOS_T2g@1i0}(b)].
Though the SDOS of the spin-3/2 $T_{2g}(0,i,1)$ state also has a ZEP, it becomes less pronounced or blunted.
This blunting effect is a consequence of the appearance of the BFSs:
The peak width is determined by the dispersion shift, being in proportion to $(\Delta_0/\mu)^2$.
Here, each of SDOS of the spin-3/2 $T_{2g}(0,i,1)$ state has several small peaks, which is attributed to the contribution of the surface state that originally exists in the normal state (see the discussion around Fig.~\ref{fig:snap_T2g@1ee}).

\begin{figure}[htbp]
  \begin{center}
    \includegraphics[width=\linewidth]{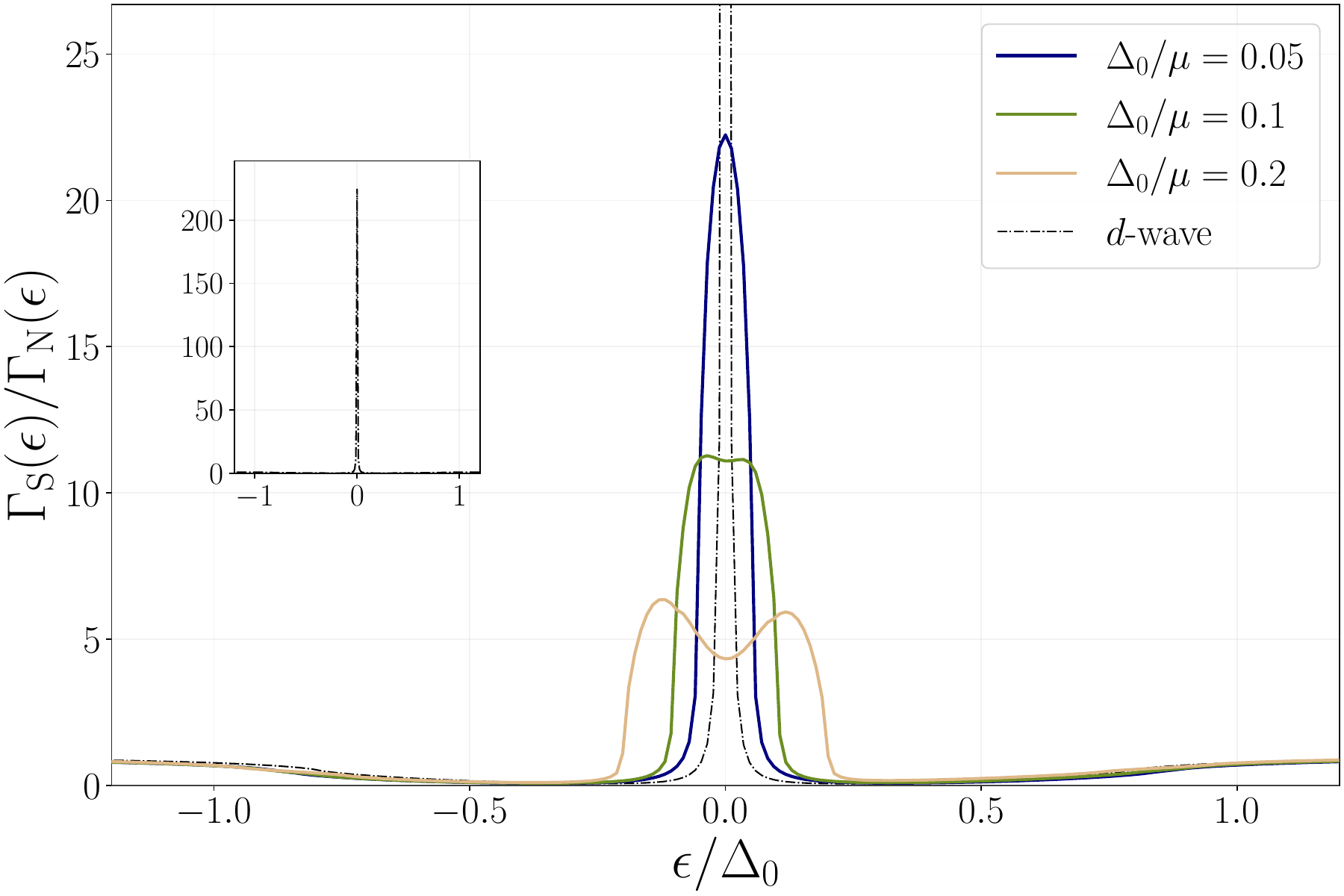}
    \caption{
        Normalized charge conductance $\Gamma_\textrm{S}(\epsilon)/\Gamma_\textrm{N}(\epsilon)$ at N/S junction with $Z=10$ for 3D chiral \textit{d}-wave and spin-3/2 $T_{2g}(0,i,1)$ states,
        where $\Gamma_\textrm{N}(\epsilon)$ is calculated by using the same code as $\Gamma_\textrm{S}(\epsilon)$ but with $\Delta_0=0$.
        Inset is the result for 3D chiral \textit{d}-wave state with an expanded vertical axis, which is independent of the value of $\Delta_0/\mu.$
        In accordance with the change in SDOS in Fig.~\ref{fig:LDOS_T2g@1i0}(b), the appearance of the BFSs blunts the ZBCP.
    }
    \label{fig:COND_T2g@1i0}
  \end{center}
\end{figure}
Figure~\ref{fig:COND_T2g@1i0} displays the results of the normalized charge conductance $\Gamma_\textrm{S}(\epsilon)/\Gamma_\textrm{N}(\epsilon)$ in the N/S junction with barrier parameter $Z=10$.
In the case of the 3D chiral \textit{d}-wave superconductor, the ZBCP associated with the ZEP in SDOS emerges as known in the previous studies~\cite{Kobayashi2015,Tamura2017,SuzukiGolubov2023}.
The charge conductance for the spin-3/2 $T_{2g}$(0,i,1) state also reflects the structure in SDOS: Similarly to the ZEP in Fig.~\ref{fig:LDOS_T2g@1i0}(b), the ZBCP is blunted as a consequence of the appearance of the BFSs.
This blunting of the ZBCP is an observable signature of the BFSs.

\begin{figure}[htbp]
  \begin{center}
    \includegraphics[width=\linewidth]{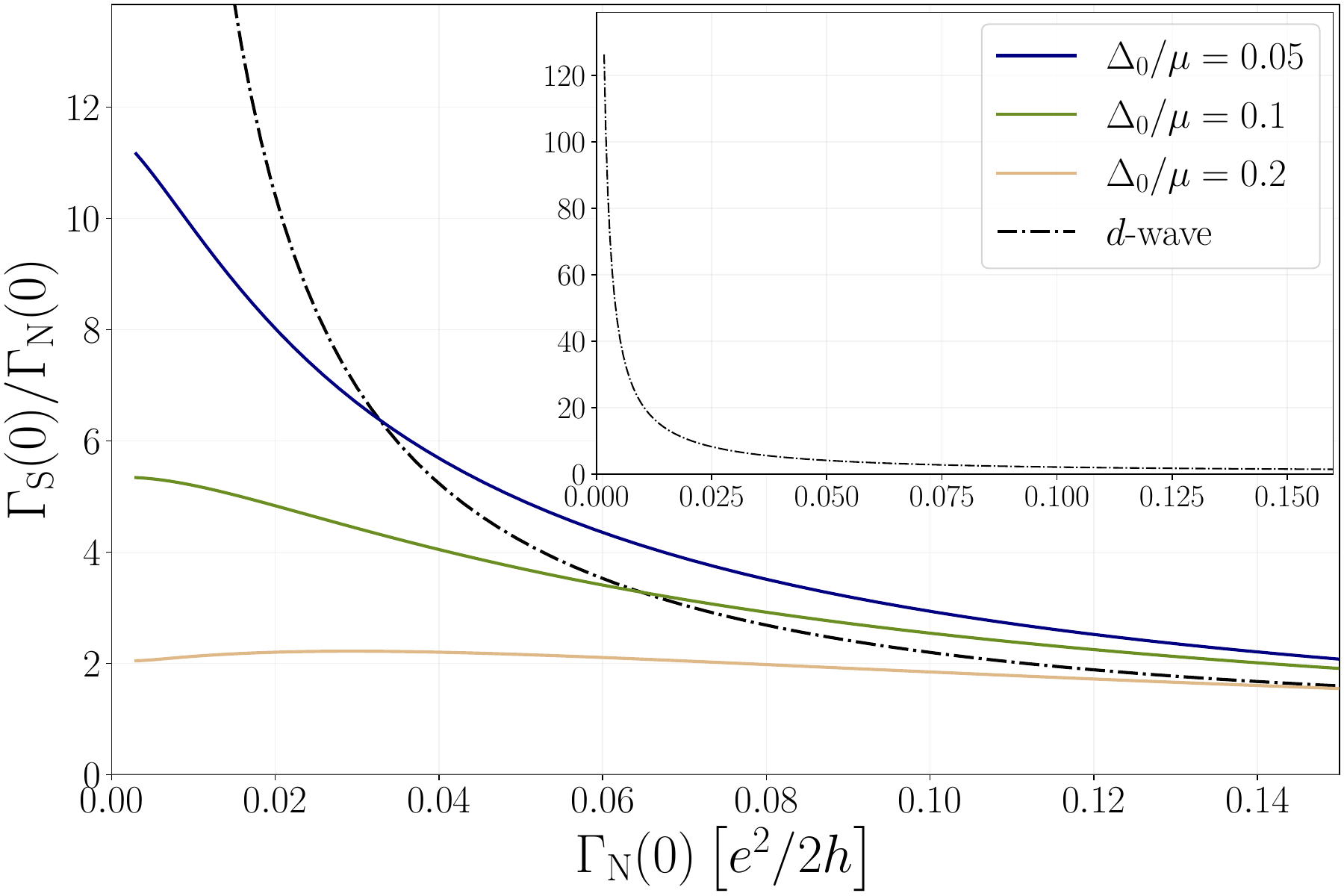}
    \caption{
        The transmittance dependence of ZBCP for 3D chiral \textit{d}-wave and spin-3/2 $T_{2g}(0,i,1)$ states at $Z=10$. 
        Inset is the result in 3D chiral \textit{d}-wave SC state with an expanded vertical axis, which is independent of the value of $\Delta_0/\mu$.
        In the case of spin-3/2 $T_{2g}(0,i,1)$ state, the ZBCP is more suppressed for a larger $\Delta_0/\mu$ and remains finite even at low transmittance.
    }
    \label{fig:COND_vs_barrier_T2g@1i0}
  \end{center}
\end{figure}
At the end of this subsection, we discuss the normalized charge conductance at zero energy.
Figure~\ref{fig:COND_vs_barrier_T2g@1i0} shows the relationship between the normal conductance and the normalized charge conductance.
It is known that the zero-energy conductance of the 3D chiral \textit{d}-wave superconductor, regardless of the magnitude of the pair amplitude, diverges as the normal state conductance decreases (or barrier parameter $Z$ increases), according to the relation $\Gamma_\mathrm{S}/\Gamma_\mathrm{N}\propto 1/\Gamma_\mathrm{N}$~\cite{TK95}.
By contrast, the zero-energy conductance of the spin-3/2 $T_{2g}(0,i,1)$ state takes a finite value in the limit of $\Gamma_\mathrm{N}\to 0$, which decreases as $\Delta_0/\mu$ increases.
This result can be explained by noting the fact that
$\Gamma_\textrm{S}(\epsilon=0)$ in the presence of BFSs consists of two contributions:
One is the Andreev reflection via the zero-energy SABSs;
The other is the residual electrical conduction via the zero-energy states on the BFSs.
In the present case, the former contribution is strongly suppressed because most of the SABSs shift from zero energy due to the existence of the BFSs [see the difference between Figs.~\ref{fig:SDOS_T2g@1i0_zero_ene}(a) and \ref{fig:SDOS_T2g@1i0_zero_ene}(b)(c)]~\cite{Kobayashi2015}.
On the other hand, the latter behaves similarly to the normal-state conduction and goes zero as $\Gamma_\textrm{N}\to 0$.
The combined effect of these two results in a finite value of $\Gamma_\textrm{S}/\Gamma_\textrm{N}$ at $\Gamma_\textrm{N}\to 0$.

\subsection{\texorpdfstring{$\bm{T_{2g}\;(1,\omega,\omega^2)}$}{}}
\textit{Short summary for $T_{2g}(1,\omega,\omega^2)$ state}:
In both the 3D cyclic \textit{d}-wave state and the spin-3/2 $T_{2g}(1,\omega,\omega^2)$ state, topological surface arc states arise because the point nodes and BFSs projected onto the surface Brillouin zone have nonzero Chern numbers.
In the 3D cyclic \textit{d}-wave state, the location of the arcs is topologically restricted on the $k_x=0$, $k_y=0$, and $k_x=k_y$ lines, and hence, the arcs cross at $k_x=k_y=0$, whereas such a constraint is removed in the spin-3/2 $T_{2g}(1,\omega,\omega^2)$ state, resulting in the disappearance of the crossing point of the arcs.
This change splits the ZEP in SDOS and the ZBCP in the 3D cyclic \textit{d}-wave state and two small peaks appear at nonzero energies in the spin-3/2 $T_{2g}(1,\omega,\omega^2)$ state.

\begin{figure}[htbp]
  \centering
  \includegraphics[width=\linewidth]{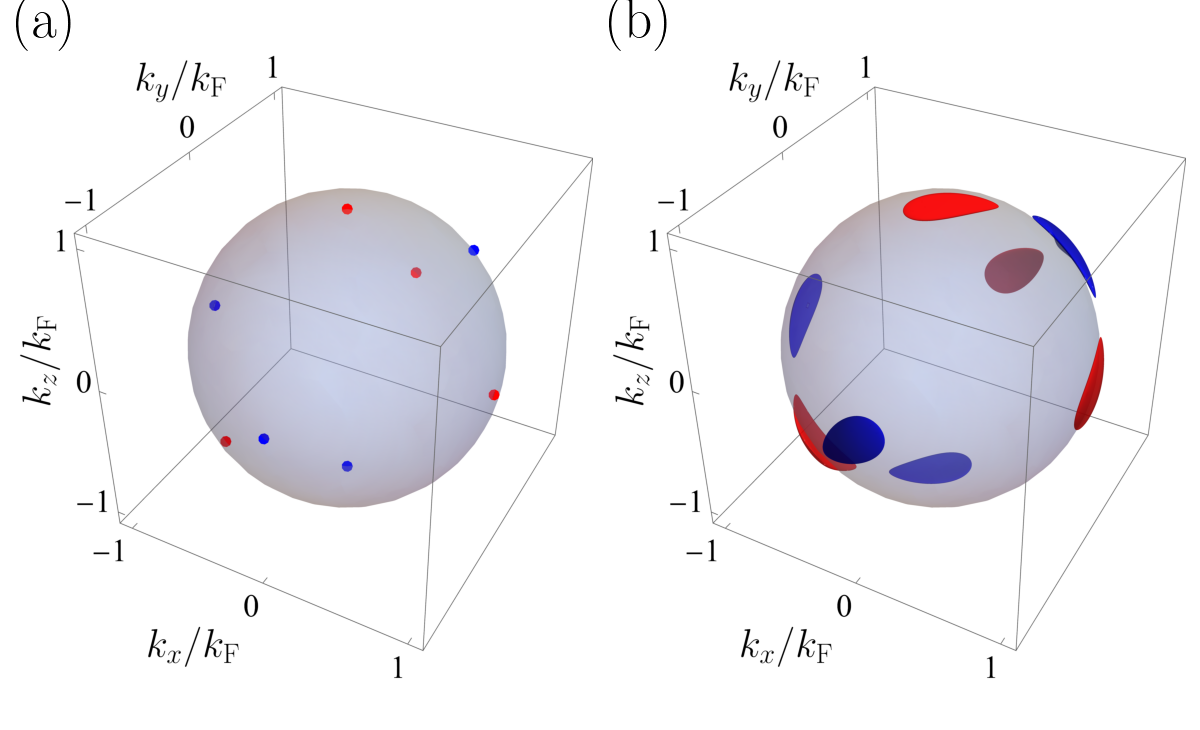}
  \caption{
      (a) Point nodes in 3D cyclic \textit{d}-wave superconducting state [Eq.~\eqref{eq:3states_spin1/2_b}] and (b) BFSs in spin-3/2 $T_{2g}(1,\omega,\omega^{2})$ state [Eq.~\eqref{eq:3states_spin3/2_b}] with $\Delta_{0}/\mu=0.2$.
      Red (blue) color on the point nodes and BFSs indicates that the Chern number defined on a surface enclosing the point node or the BFS is $+2$ ($-2$).
  }
  \label{fig:BFS_T2g@1ee}
\end{figure}

The pair potentials for the $T_{2g}(1,\omega,\omega^2)$ states of $j=1/2$ and $j=3/2$ systems are 
given by Eqs.~\eqref{eq:3states_spin3/2_b} and \eqref{eq:3states_spin1/2_b}, respectively.
The spin-1/2 pairing state is the 3D cyclic \textit{d}-wave superconducting state, which has a pair of point nodes along each of the $k_x=k_y=k_z$ line and the $x$, $y$, and $z$ axes, and has eight point nodes in total [Fig.~\ref{fig:BFS_T2g@1ee}(a)].
Correspondingly, the spin-3/2 $T_{2g}(1,\omega,\omega^2)$ state has eight BFSs extend alongside the Fermi surface as illustrated in Fig.~\ref{fig:BFS_T2g@1ee}(b), which is calculated by using Eq.~\eqref{eq:Pf_gen}.
The color of each node and each BFS in Fig.~\ref{fig:BFS_T2g@1ee} indicates the sign of Chern number $\Ch$ at each node, where the red (blue) color denotes $\Ch=+2 (-2)$. Here, Chern number
is defined by~\cite{Brydon2018,Kobayashi2015}:
\begin{align}
   \Ch
   = \frac{i}{2\pi}\sum_{n\in  \mathrm{occ}} \oint _{\mathcal{S}} d^{2}\bm{s}(\kk)\cdot \left[\nabla_{\kk}\times \braket{u_{n}(\kk)|\nabla_{\kk}|u_{n}(\kk)}\right],
   \label{eq:Chern_number}
\end{align}
where $\mathcal{S}$ is a closed surface surrounding the node or the BFS, $d^{2}\bm{s}(\kk)$ is a vectorial surface element in the momentum space, $\ket{u_{n}(\kk)}$ is an eigenstate of the BdG Hamiltonian $H_\kk$, and the summation is taken over all of the occupied states.
The BFSs spread alongside the Fermi surface with the increase of $\Delta_0$, and eventually, four of them merge into one when $\Delta_0$ exceeds a critical value ($\Delta_{0}/\mu \sim 0.2168$ for our setup). 
The Chern number for the combined BFS is the summation of those for each BFS before the merge.

\begin{figure}[htbp]
  \begin{center}
    \includegraphics[width=\linewidth]{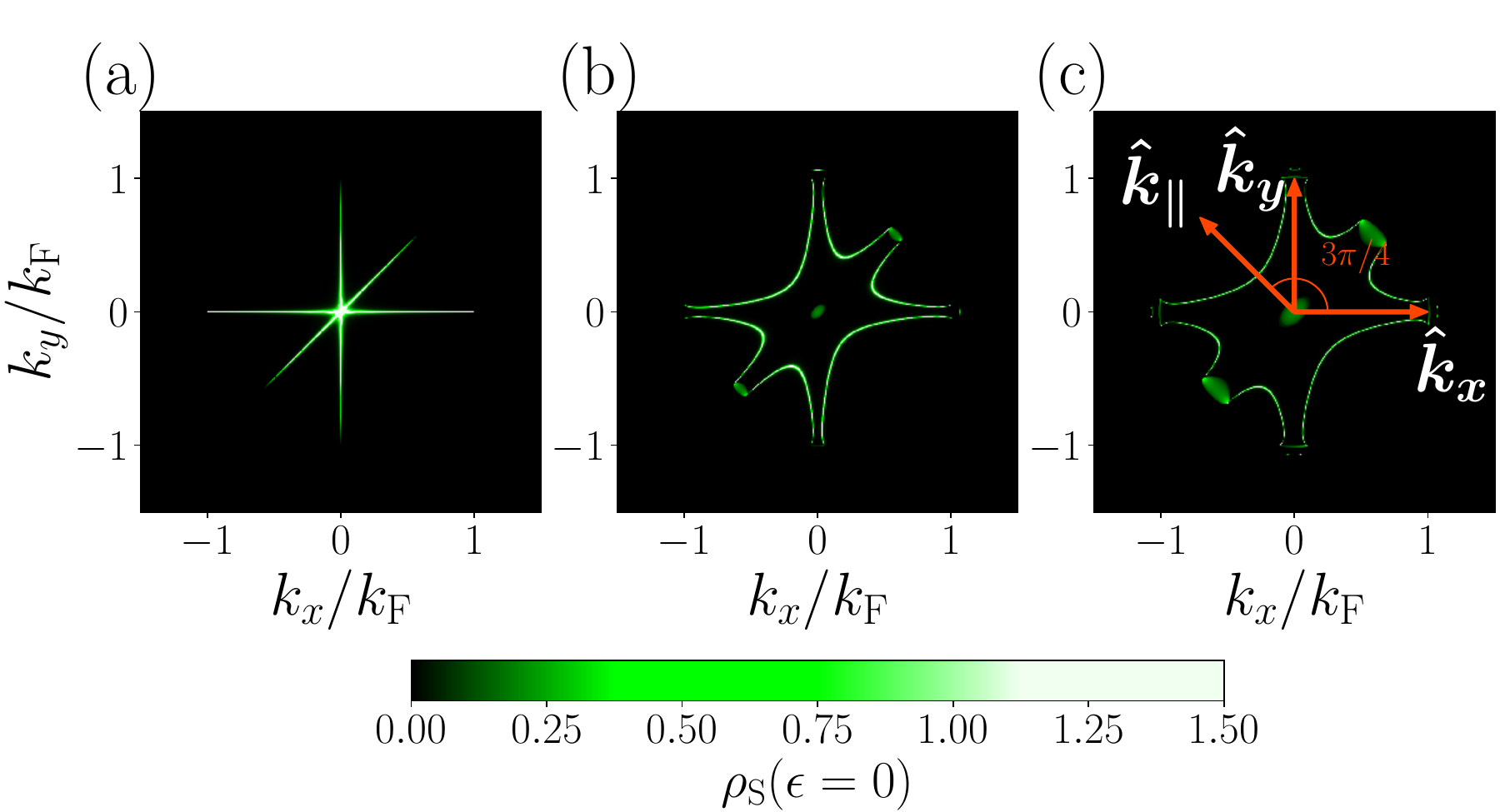}
    \caption{
        Momentum-resolved zero-energy SDOS for 3D cyclic \textit{d}-wave (a) and spin-3/2 $T_{2g} (1,\omega,\omega^2)$ states with $\Delta_0/\mu=0.05$ (b) and $0.1$ (c).
        Panel (a) is independent of the value of $\Delta_0/\mu$.
        Two arcs come out of each of the point nodes (a) or BFSs (b), (c) projected onto the surface Brillouin zone. 
        The BFSs split the doubly-degenerate arc states in (a).
        }
    \label{fig:SDOS_T2g@1ee_zero_ene}
  \end{center}
\end{figure}
\begin{figure}[htbp]
  \begin{center}
    \includegraphics[width=\linewidth]{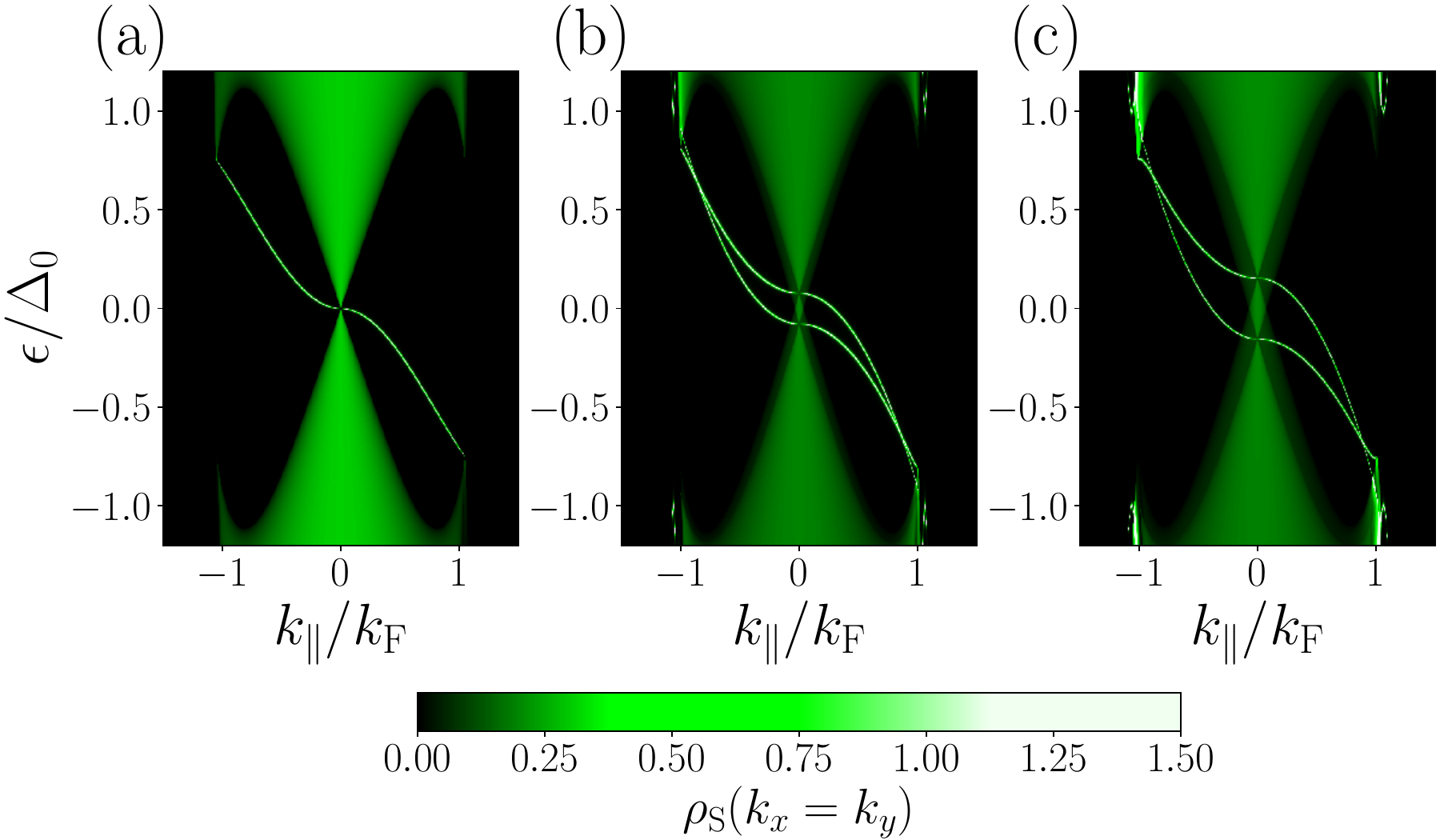}
  \caption{
    Momentum-resolved SDOS along the $\hat{k}_\parallel$ direction shown in Fig.~\ref{fig:SDOS_T2g@1ee_zero_ene}(c) for spin-1/2 $T_{2g}(1,\omega,\omega^2)$ state (a) and spin-3/2 $T_{2g} (1,\omega,\omega^2)$ states with $\Delta_0/\mu=0.05$ (b), and $0.2$ (c). 
    Panel (a) is independent of the value of $\Delta_0/\mu$.
    }
    \label{fig:SDOS_T2g@1ee}
  \end{center}
\end{figure}
Figure~\ref{fig:SDOS_T2g@1ee_zero_ene} and \ref{fig:SDOS_T2g@1ee} illustrate the momentum-resolved SDOS at $\epsilon=0$ and at $k_y = -k_x$ for the 3D cyclic \textit{d}-wave superconductor (a), and for the spin-3/2 $T_{2g}(1,\omega,\omega^2)$ superconductor at $\Delta_0/\mu=0.05$ (b) and $0.1$ (c).
In Fig.~\ref{fig:SDOS_T2g@1ee_zero_ene}(a), arc states arise on the $k_x=0$, $k_y=0$, and $k_x=k_y$ lines, connecting the nodal points projected onto the surface Brillouin zone. 
The location of the arcs is protected by the 1D winding number associated with the chiral symmetry: 
In addition to the inversion symmetry, the Hamiltonian with pair potential Eq.~\eqref{eq:3states_spin1/2_b} has the pseudo-time-reversal symmetry on these lines, i.e., $H_{\bm k}$ satisfies $\tilde{T}^\dagger H_{\kk}\tilde{T}=H_{-\kk}$ where $\tilde{T}=U^\dagger T U$ with $T=i\sigma_y\tau_0 \mathcal{K}$ being the conventional time-reversal operator and $U=e^{-i(\pi/3)\sigma_0\tau_z}, e^{-i(2\pi/3)\sigma_0\tau_z}$ and $1$ for $H_{\bm k}$ on the lines $k_x=0, k_y=0$, and $k_x=k_y$, respectively~\cite{Kobayashi2015}.
The winding number further reveals that the arc states on these lines are doubly degenerate.
In the spin-3/2 $T_{2g}(1, \omega, \omega^2)$ state, however, the pseudo-time-reversal symmetry on these lines is no longer preserved, and the location of the arcs changes. 
Nevertheless, the total number of arcs does not change because each of the point nodes or BFSs projected onto the surface Brillouin zone is attached to two arcs, reflecting their winding number of $|\Ch|=2$. 
The doubly-degenerate arc states in the 3D cyclic \textit{d}-wave state split into two, and the splitting becomes larger as $\Delta_0$ increases.

\begin{figure}[htbp]
  \begin{center}
    \includegraphics[width=\linewidth]{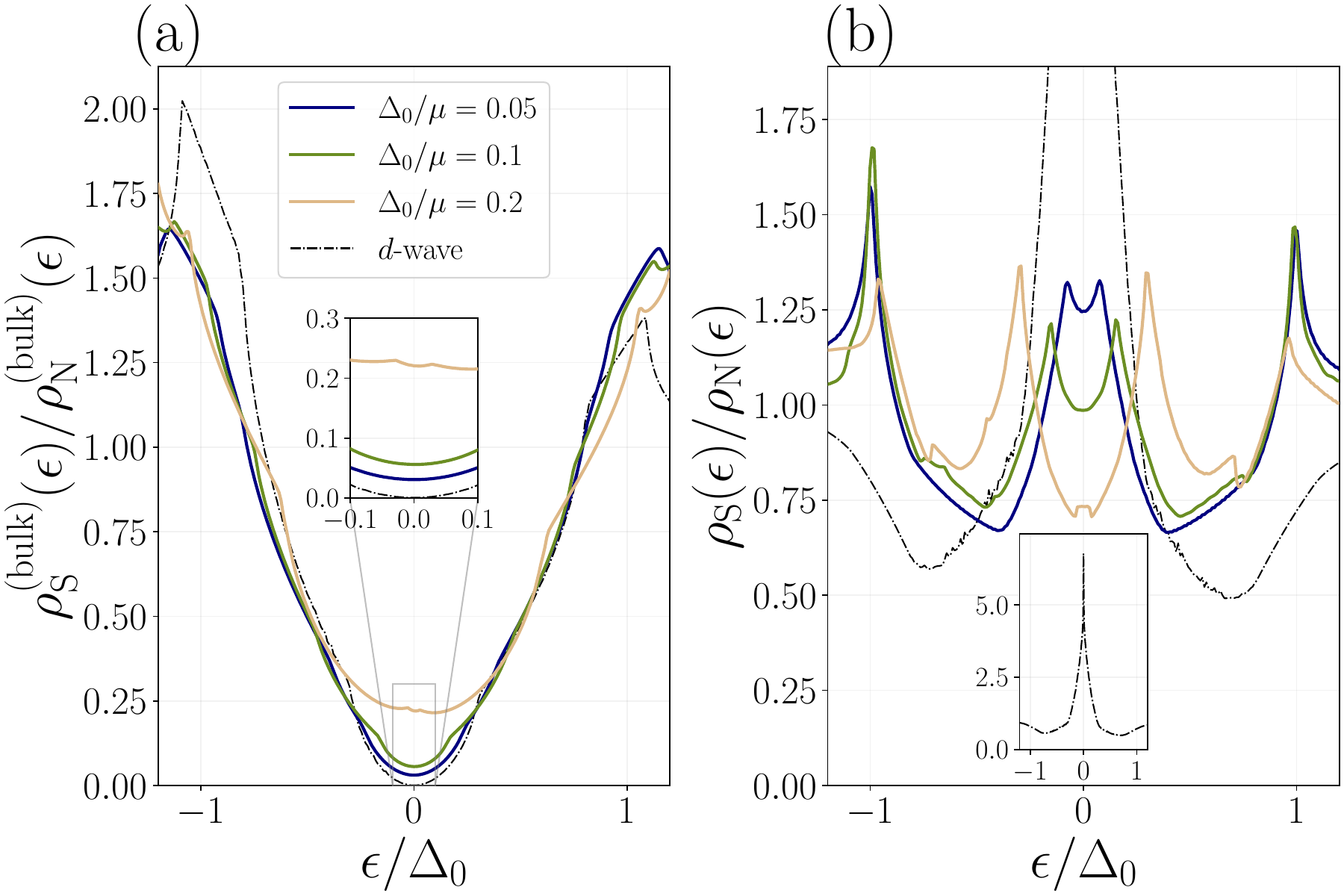}
    \caption{
    Bulk LDOS (a) and SDOS (b) of $T_{2g}(1,\omega,\omega^2)$ states normalized by their values at $\Delta_0=0$. 
    Inset in (a) is a magnified view around $\epsilon=0$, and inset in (b) is the SDOS of 3D cyclic \textit{d}-wave state with an expanded vertical axis.
    The results for the \textit{d}-wave state are independent of the value of $\Delta_0/\mu.$
    (a) The BFSs contribute to the nonzero bulk LDOS at $\epsilon=0$.
    (b) The appearance of the BFSs splits the ZEP in the \textit{d}-wave case.
    }
    \label{fig:LDOS_T2g@1ee}
  \end{center}
\end{figure}
\begin{figure}[htbp]
  \begin{center}
    \includegraphics[width=\linewidth]{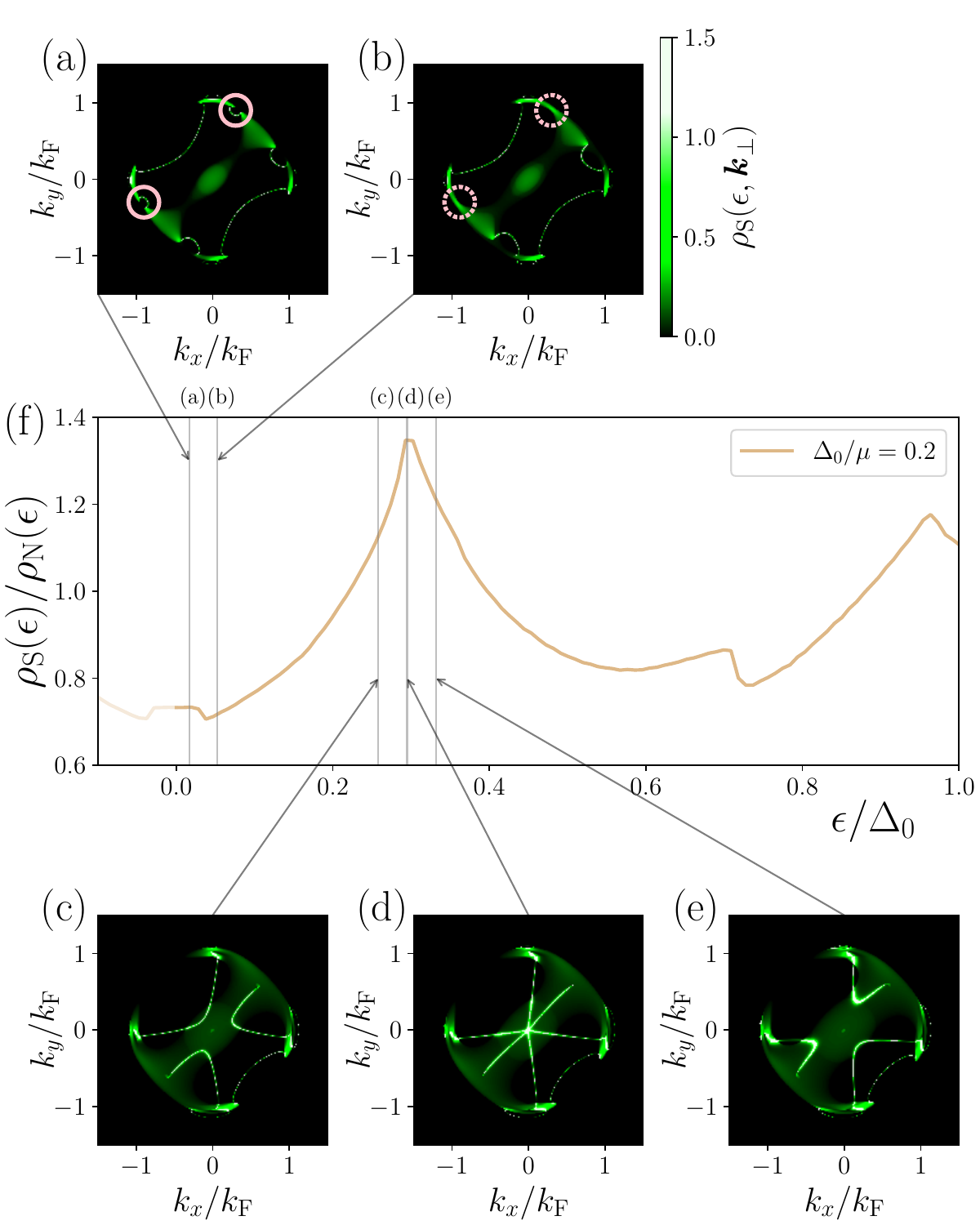}
    \caption{
        Relationship between SABSs and SDOS in spin-3/2 $T_{2g}(1,\omega,\omega^2)$ state.
        (a)-(e) Momentum-resolved SDOS at different energy levels indicated by arrows.
        (f) SDOS of spin-3/2 $T_{2g}(1,\omega,\omega^2)$ state at $\Delta_0/\mu=0.2$.
        The data in (f) is the same as that shown in Fig.~\ref{fig:LDOS_T2g@1ee}(b).
    }
      \label{fig:snap_T2g@1ee}
  \end{center}
\end{figure}

We show the bulk LDOS and the SDOS in Figs.~\ref{fig:LDOS_T2g@1ee}(a) and \ref{fig:LDOS_T2g@1ee}(b), respectively.
In the case of the 3D cyclic \textit{d}-wave superconductor with the point nodes shown in Fig.~\ref{fig:BFS_T2g@1ee}(a), the bulk LDOS exhibits a characteristic U-shaped structure, with a zero value at zero energy~\cite{Ishikawa2013}.
The bulk LDOS in the spin-3/2 $T_{2g}(1,\omega,\omega^2)$ state also has the same U-shaped structure. 
However, as in the case of $T_{2g}(0, i, 1)$ state, there are nonzero $\rho_\mathrm{S}^\textrm{(bulk)}$ values at zero energy.
This increment at zero energy is nothing but the contributions from the states on the BFSs.
On the other hand, more drastic change happens in the SDOS:
Whereas the SDOS of the 3D cyclic \textit{d}-wave state has a ZEP with finite width, that of the spin-3/2 $T_{2g}(1\omega,\omega^2)$ state exhibits distinct dip and peak structures.
The disappearance of the ZEP is the consequence of the vanishing crossing point of arcs at $k_x=k_y=0$ in Fig.~\ref{fig:SDOS_T2g@1ee_zero_ene}(a).

We can further understand the dip and peak structures of the SDOS by analyzing the momentum-resolved SDOS with changing $\epsilon$.
In Fig.~\ref{fig:snap_T2g@1ee}, we show the magnified view of the result for $\Delta_0/\mu=0.2$ in Fig.~\ref{fig:LDOS_T2g@1ee}(b) together with the momentum-resolved SDOS at each $\epsilon$ indicated by arrows.
By comparing the SDOS in Figs.~\ref{fig:snap_T2g@1ee}(a) and \ref{fig:snap_T2g@1ee}(b), in particular inside the pink solid circles and pink dashed circles, one can see that the small dip at $\epsilon/\mu\sim 0.04$ in Fig.~\ref{fig:snap_T2g@1ee}(f) is associated with the vanishing of two SABSs encircled by pink solid curves in Fig.~\ref{fig:snap_T2g@1ee}(a).
On the other hand, the peak at $\epsilon/\mu\sim 0.3$ in Fig.~\ref{fig:snap_T2g@1ee}(f) is due to the reconnection of the SABSs as seen in Figs.~\ref{fig:snap_T2g@1ee}(c)-(e):
In the course of the reconnection, the SABSs have a crossed structure [Fig.~\ref{fig:snap_T2g@1ee}(d)] as in the 3D cyclic \textit{d}-wave case [Fig.~\ref{fig:SDOS_T2g@1ee_zero_ene}(a)], leading to a peak in the SDOS.
In general, a crossing of surface states leads to a peak in SDOS~\cite{LuBo}.
In our model, the normal-state Hamiltonian has a topological surface state with sharp dispersion, whose contribution to the momentum-resolved SDOS is too narrow to be recognized.
The small peaks of SDOS in Figs.~\ref{fig:LDOS_T2g@1i0}(b) and \ref{fig:LDOS_Eg@1i}(b) are attributed to the reconnection between SABSs and the surface state in normal state.

\begin{figure}[htbp]
  \begin{center}
    \includegraphics[width=\linewidth]{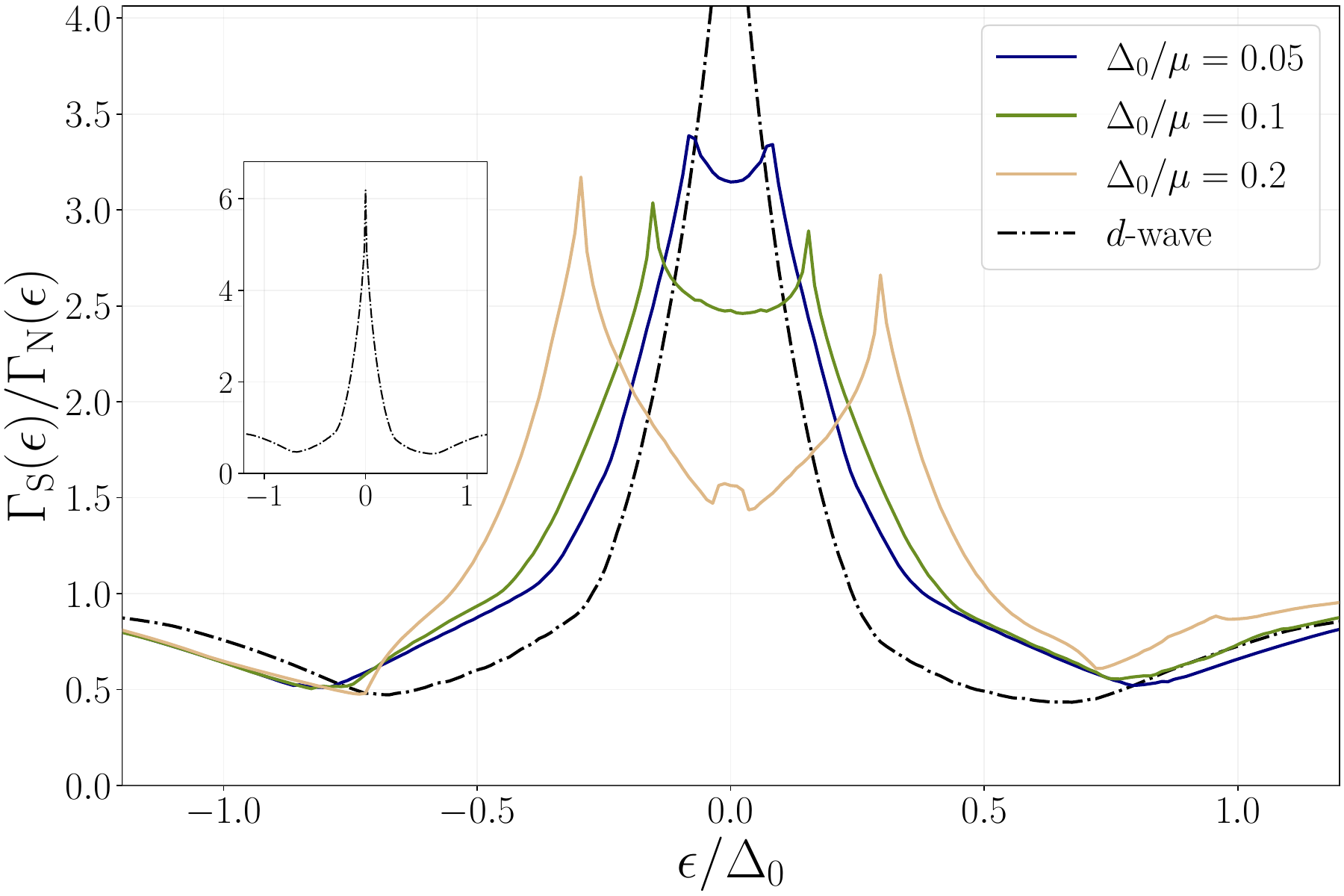}
    \caption{
        Normalized charge conductance $\Gamma_\textrm{S}(\epsilon)/\Gamma_\textrm{N}(\epsilon)$ at N/S junction with $Z=10$ for 3D cyclic \textit{d}-wave and spin-3/2 $T_{2g}(1,\omega,\omega^{2})$ states.
        Inset is the result for 3D chiral \textit{d}-wave state with an expanded vertical axis, which is independent of the value of $\Delta_0/\mu.$
        In accordance with the change in SDOS in Fig.~\ref{fig:LDOS_T2g@1ee}(b), the appearance of the BFSs splits the ZBCP.
    }
    \label{fig:COND_T2g@1ee}
  \end{center}
\end{figure}

Figure~\ref{fig:COND_T2g@1ee} shows the normalized charge conductance at the N/S junction with the barrier parameter $Z=10$. 
In both cases of the 3D cyclic \textit{d}-wave state and the spin-3/2 $T_{2g}(1,\omega,\omega^2)$ state, the normalized charge conductance has a similar structure as the SDOS in Fig.~\ref{fig:LDOS_T2g@1ee}(b).
Namely, the ZBCP in the 3D cyclic \textit{d}-wave state splits into two small peaks at nonzero energy with the appearance of the BFSs.

\subsection{\texorpdfstring{$\bm{E_{g}\;(1,i)}$}{}}\label{ssec:Eg(1,i)}
\textit{Short summary for $E_g(1,i)$ state}:
In the absence of the topological SABSs, no peak structure appears around zero energy in the SDOS nor the charge conductance.
In accordance with the deformation from point nodes to BFSs, nonzero SDOS arises at $\epsilon=0$, which leads to a nonzero electron conduction in the N/S junction at zero bias.

\begin{figure}[htbp]
  \centering
\includegraphics[width=\linewidth]{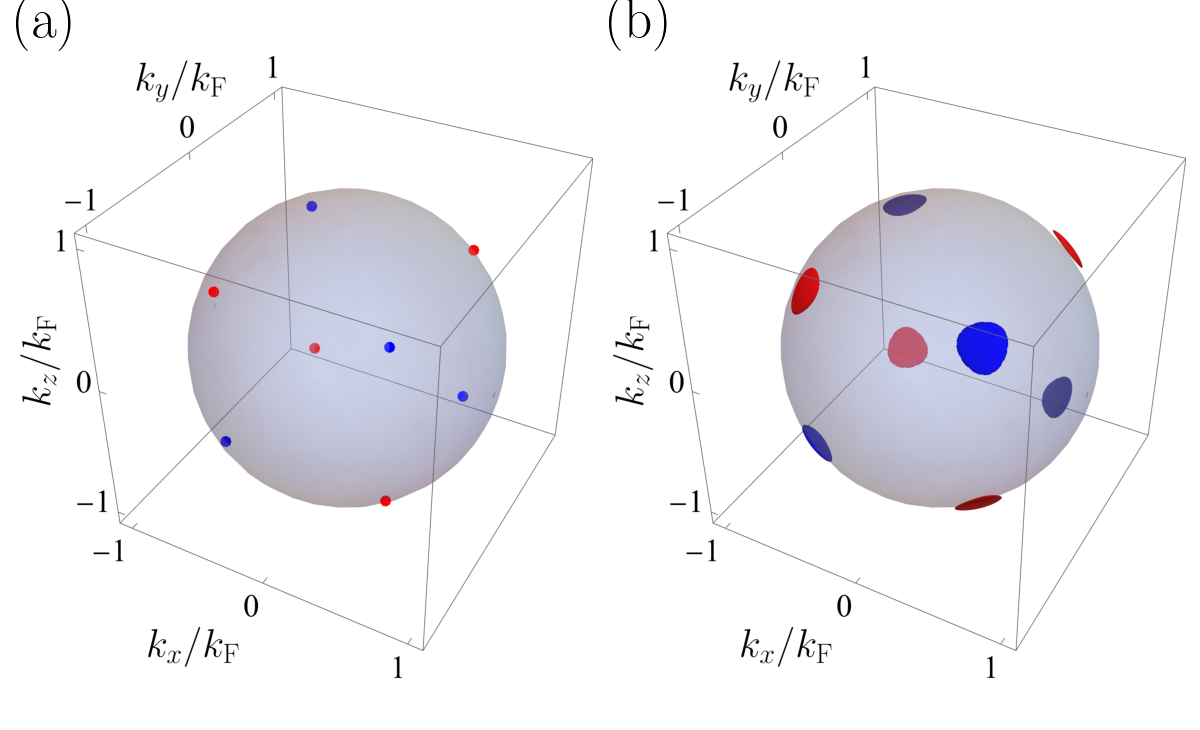}
  \caption{
      (a) Point nodes in spin-1/2 $E_g(1,i)$ state [Eq.~\eqref{eq:3states_spin1/2_c}] and (b) BFSs in spin-3/2 $E_{g}(1,i)$ state [Eq.~\eqref{eq:3states_spin3/2_c}] with $\Delta_{0}/\mu=0.2$.
      Red (blue) color on the point nodes and BFSs indicates that the Chern number defined on a surface enclosing the point node or the BFS is $+2$ ($-2$).
  }
  \label{fig:BFS_Eg@1i}
\end{figure}

As a third case, we consider the $E_{g}(1,i)$ state, whose pair potential is given by Eqs.~\eqref{eq:3states_spin1/2_c} and \eqref{eq:3states_spin3/2_c} for spin-$1/2$ and spin-$3/2$ superconductors, respectively.
The spin-1/2 $E_g(1,i)$ state has eight point nodes along the $(\pm1,\pm1,\pm1)$ directions as shown in Fig.~\ref{fig:BFS_Eg@1i}(a).
Correspondingly, the spin-3/2 $E_g(1,i)$ state has eight BFSs that extend alongside the Fermi surface as illustrated in Fig.~\ref{fig:BFS_Eg@1i}(b).
As in the case of the $T_{2g}(1,\omega,\omega^2)$ state, the point nodes and BFSs in the $E_g(1,i)$ state have nonzero Chern numbers.
The red (blue) color on the nodes and BFSs indicates that the Chern number [Eq.~~\eqref{eq:Chern_number}] of the node and the BFS is $+2$ ($-2$). 
\begin{figure}[htbp]
  \begin{center}
    \includegraphics[width=\linewidth]{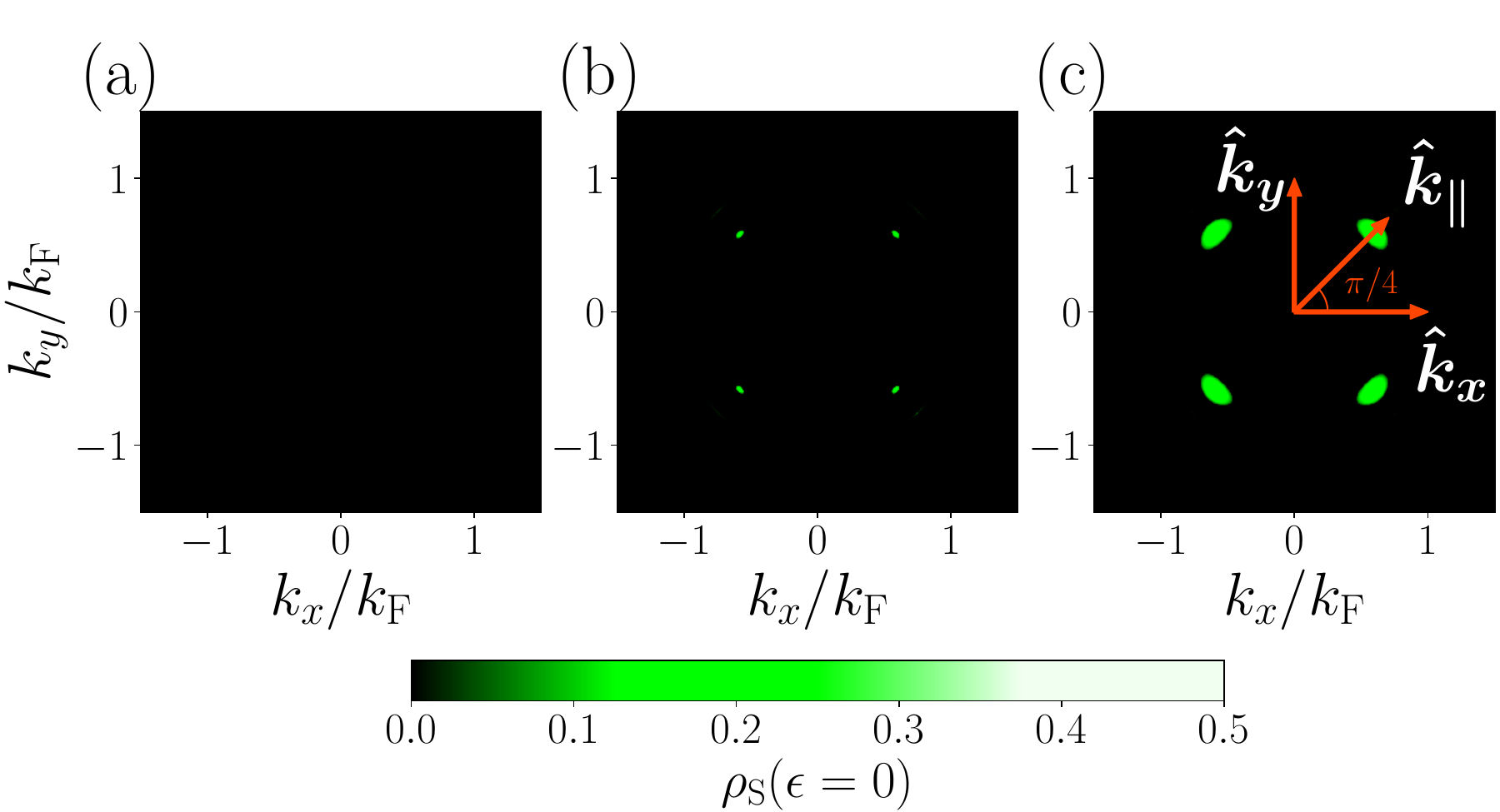}
    \caption{
        Momentum-resolved zero-energy SDOS for spin-1/2 $E_g(1,i)$ state (a) and spin-3/2 $E_g(1,i)$ state with $\Delta_0/\mu=0.05$ (b) and $0.2$ (c).
        Panel (a) is independent of the value of $\Delta_0/\mu$. 
        In (a), $\rho_\textrm{S}(0)$ is nonzero only on the point nodes, which are invisible in this graph. 
        The area of nonzero $\rho_\textrm{S}(0)$ increases as $\Delta_0$ increases in (b) and (c).
        In all cases, no SABS appears.
        }
    \label{fig:SDOS_Eg@1i_zero_ene}
  \end{center}
\end{figure}

\begin{figure}[htbp]
  \begin{center}
    \includegraphics[width=\linewidth]{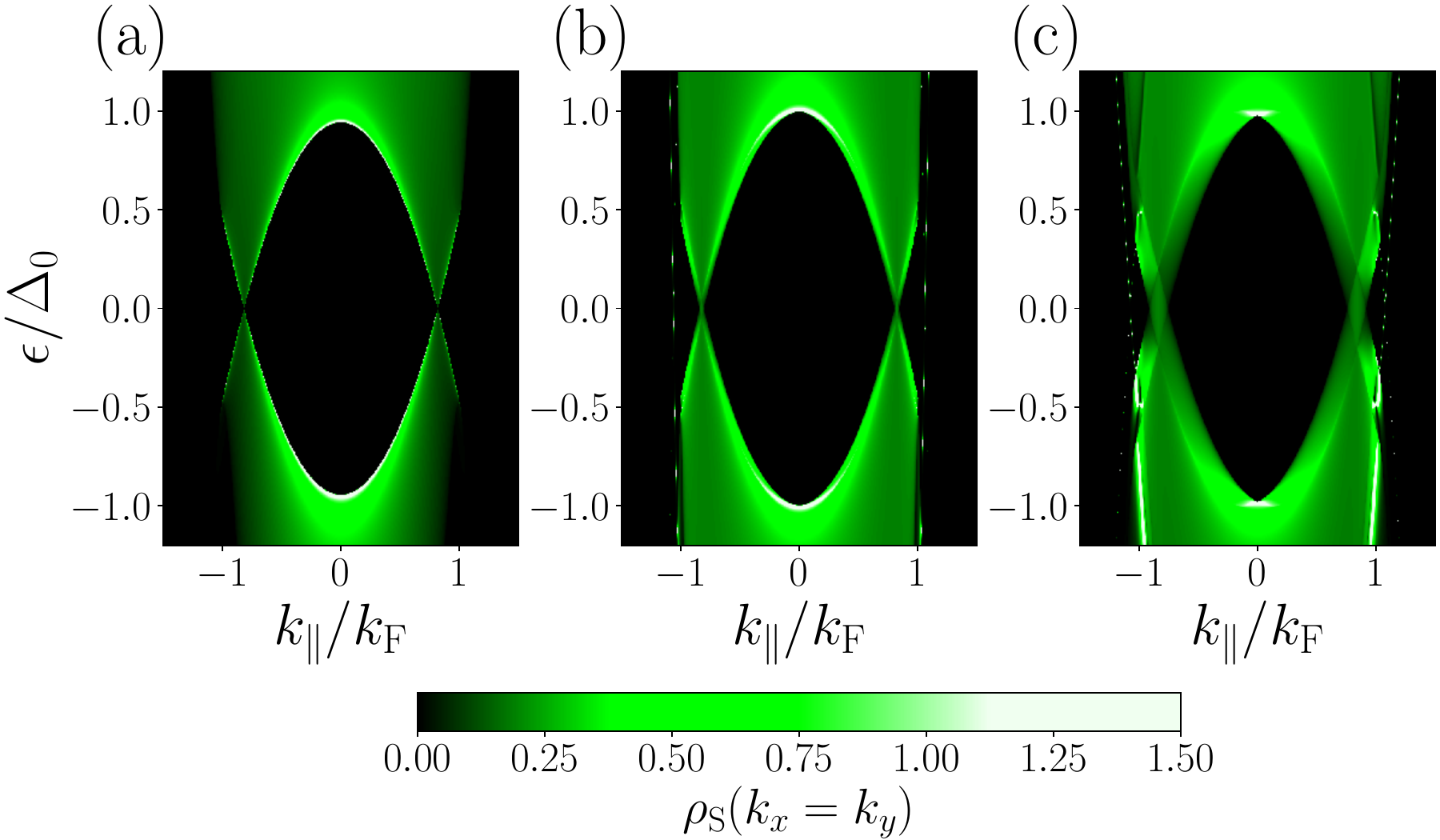}
  \caption{
    Momentum-resolved SDOS along the $\hat{k}_\parallel$ direction shown in Fig.~\ref{fig:SDOS_Eg@1i_zero_ene}(c) for spin-1/2 $E_g(1,i)$ state (a) and spin-3/2 $E_{g} (1,i)$ states with $\Delta_0/\mu=0.05$ (b), and $0.2$ (c). 
    Panel (a) is independent of the value of $\Delta_0/\mu$.
    }
    \label{fig:SDOS_Eg@1i}
  \end{center}
\end{figure}

We show the momentum-resolved SDOS at $\epsilon=0$ and at $k_x=k_y$ in Figs.~\ref{fig:SDOS_Eg@1i_zero_ene} and \ref{fig:SDOS_Eg@1i}, respectively.
One can see from Fig.~\ref{fig:SDOS_Eg@1i_zero_ene} that the BFSs expand as $\Delta_0$ increases.
However, no SABS is observed on the (001) surface.
The absence of SABS is also confirmed from the momentum-resolved SDOS on the $k_\parallel-\epsilon$ space shown Fig.~\ref{fig:SDOS_Eg@1i}, where $k_\parallel$ is the momentum along the $\hat{k}_\parallel$ direction depicted in Fig.~\ref{fig:SDOS_Eg@1i_zero_ene}(c).
The absence of SABS is because the summation of the Chern numbers of the BFSs projected onto the same place in the surface Brillouin Zone is zero.
Thus, in this case, we can investigate the transport through the zero-energy states on the BFSs without being buried by that via SABSs.
We comment that the \red{SABSs} appear when we create a surface in other directions, such as the (1,1,1) surface~\cite{Brydon2018}.

\begin{figure}[htbp]
  \begin{center}
    \includegraphics[width=\linewidth]{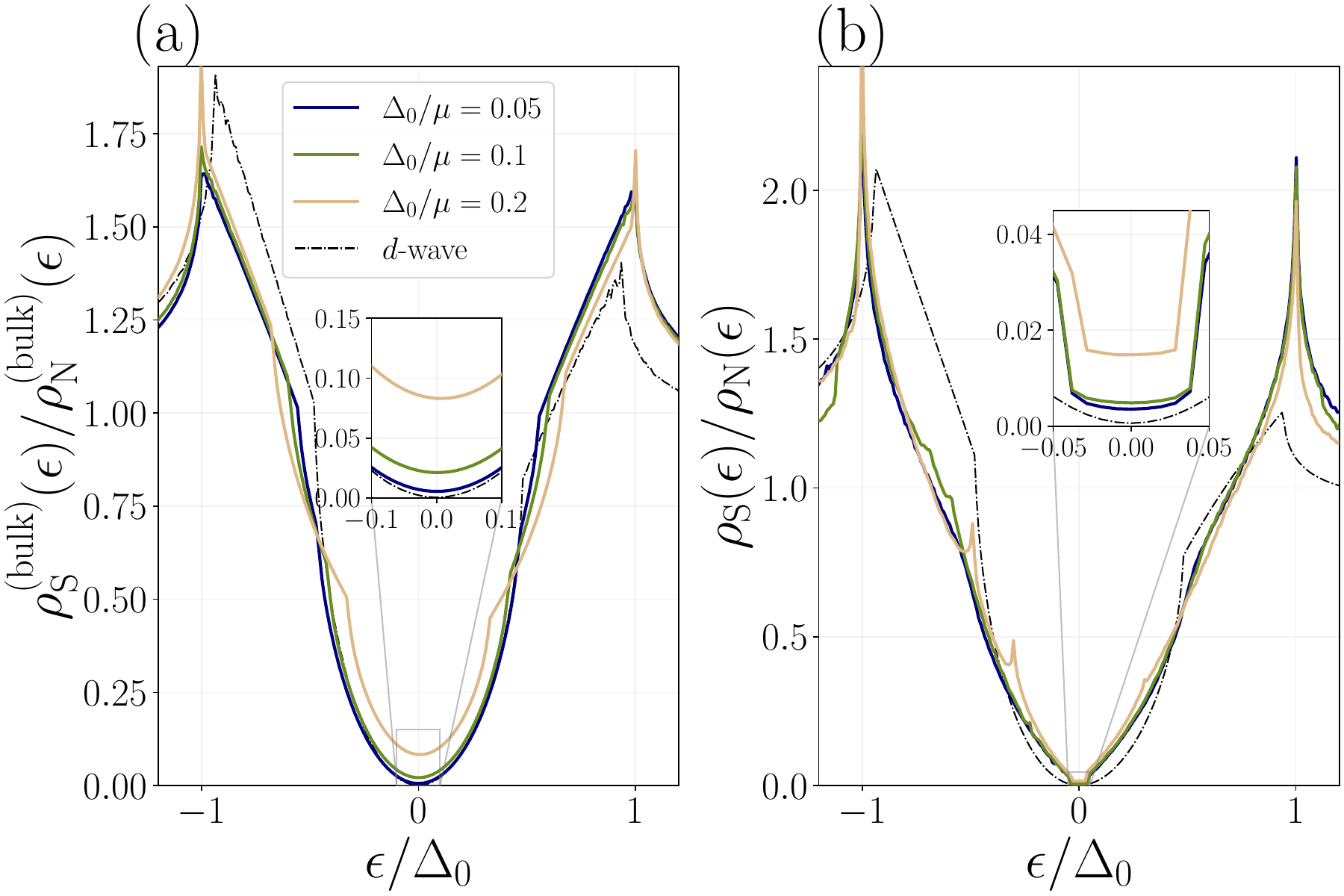}
    \caption{
        Bulk LDOS (a) and SDOS (b) of $E_{g}(1,i)$ states normalized by their values at $\Delta_0=0$.
        Insets in (a) and (b) are magnified views around $\epsilon=0$.
        The legend "\textit{d}-wave" denotes the result for the spin-1/2 $E_g(1,i)$ state, which is independent of the value of $\Delta_0/\mu$.
        There is no ZEP in SDOS because of the absence of SABSs.
        The nonzero bulk LDOS (a) and SDOS (b) at $\epsilon=0$ are the contributions from the BFSs.
   }
    \label{fig:LDOS_Eg@1i}
  \end{center}
\end{figure}

Figures~\ref{fig:LDOS_Eg@1i}(a) and \ref{fig:LDOS_Eg@1i}(b) show the bulk LDOS and the SDOS, respectively.
As in the case of $T_{2g}(1,\omega,\omega^2)$ states, the bulk LDOS has a U-shaped structure, a characteristic of superconductors with point nodes, and the value at $\epsilon=0$ increases as $\Delta_0$, or equivalently, the area of the BFSs increases.
On the other hand, because of the lack of SABSs, the SDOS has a similar structure as the bulk LDOS, differently from the case of $T_{2g}(1,\omega,\omega^2)$ states.
The residual SDOS at $\epsilon=0$ is again due to the BFSs, although its magnitude is smaller than that in bulk LDOS. 

\begin{figure}[htbp]
  \begin{center}
    \includegraphics[width=\linewidth]{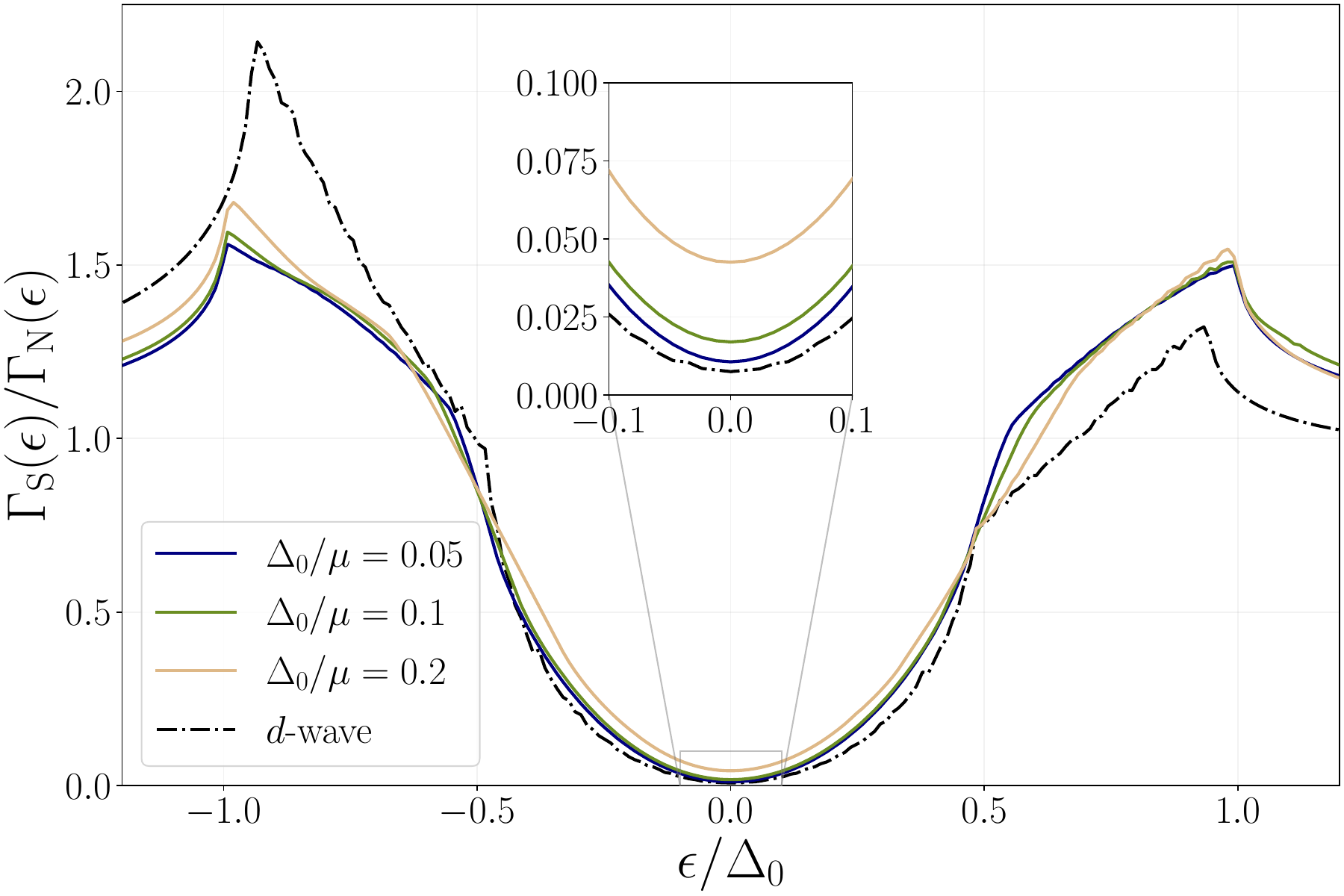}
    \caption{
        Normalized charge conductance $\Gamma_\textrm{S}(\epsilon)/\Gamma_\textrm{N}(\epsilon)$ at N/S junction with $Z=10$ for $E_{g}(1,i)$ states. 
        Inset is the magnified view around $\epsilon=0$.
        The legend "\textit{d}-wave" denotes the result for the spin-1/2 $E_{g}(1,i)$ state, 
        which is independent of the value of $\Delta_0/\mu$.
        Because of the absence of the ZBCP, the increase in the conductance at zero bias due to the transport through the states on the BFSs becomes visible.
     }
    \label{fig:COND_Eg@1i}
  \end{center}
\end{figure}

The normalized charge conductance $\Gamma_\textrm{S}(\epsilon)/\Gamma_\textrm{N}(\epsilon)$ at the N/S junction with the barrier parameter $Z=10$ is shown in Fig.~\ref{fig:COND_Eg@1i}.
Reflecting the $\epsilon$ and $\Delta_0$ dependence of the SDOS in Fig.~\ref{fig:LDOS_Eg@1i}(b), the normalized conductance has a dip instead of a peak at $\epsilon=0$.
The minimum conductance at $\epsilon=0$ in the spin-3/2 $E_g(1,i)$ state increases as $\Delta_0$ increases.
This is the residual electrical conduction via the zero-energy states on the BFSs.
We comment here that the normalized conductance at $\epsilon=0$ does not fall to zero even in the spin-1/2 $E_g(1,i)$ state due to the finite value of the barrier parameter $Z$.

\section{Conclusion}
\label{sec:conclusion}
In this paper, we have studied surface Andreev bound states (SABSs) of superconductors having Bogoliubov Fermi surfaces (BFSs) and their contributions to the charge conductance at normal metal/superconductor (N/S) junctions.
We consider spin quintet $J_\textrm{pair}=2$ pairings of spin-3/2 electrons, which exhibit anisotropic behavior due to the asymmetry in the spin space even when the pair potential is isotropic in the momentum space.
When the system is projected to an effective pseudo-spin-1/2 system, the $J_\textrm{pair}=2$ pairs are mapped to spin-singlet \textit{d}-wave pairs belonging to the same irreducible representation as the pristine $J_\textrm{pair}=2$ pair.
In particular, the $J_\textrm{pair}=2$ pairs possessing BFSs, which are our interest, are mapped to the \textit{d}-wave pair potentials with line nodes and/or point nodes, which accompany topological zero-energy surface flat bands and surface arc states on specific surfaces.
Since such topological surface states are featured by the zero-bias conductance peak (ZBCP) at an N/S junction, we have investigated how the situation changes in the spin-3/2 $J_\textrm{pair}=2$ pairs with BFSs having the same symmetry.

We have calculated the surface density of states (SDOS) and quasiparticle tunneling conductance in N/S junctions based on the recursive Green's function method and a generalized Lee-Fisher formula.
We have discussed three spin-3/2 $J_\textrm{pair}=2$ pairing states: $T_{2g}(0,i,1), T_{2g}(1,\omega,\omega^2)$, and $E_g(1,i)$ states.
The corresponding spin-1/2 \textit{d}-wave superconducting states have zero-energy surface flat bands, surface arc states, and no SABS, respectively, on the (001) surface.
In the first case, $T_{2g}(0,i,1)$, the doubly-degenerate zero-energy surface flat bands in the \textit{d}-wave pairing state split due to the appearance of the BFSs.  
Thus, the zero-energy peak (ZEP) of the SDOS and the ZBCP observed in the spin-1/2 \textit{d}-wave pairing case is blunted, although the zero-energy surface states remain on a ring on the surface Brillouin zone. 
In the second case, $T_{2g}(1,\omega,\omega^2)$, the location of the surface arc states, which is topologically restricted in the case of the spin-1/2 \textit{d}-wave pairing state, changes in the spin-3/2 system. Accordingly, the crossing of the arcs disappears, leading to the splitting of the ZEP of the SDOS and the ZBCP into two small peaks at nonzero energies.
In the third case, $E_{g}(1,i)$, there is no SABS on the (001) surface for both cases of spin-1/2 and spin-3/2 systems. 
Because there is no ZEP of the SDOS, we can investigate the electronic transport through the zero-energy states on the BFSs, and indeed, we obtained non-zero charge conductance at zero bias, which increases as the BFSs become larger.
Summarising above, the topological protections existing in the spin-1/2 \textit{d}-wave pairings are easily violated in the corresponding spin-3/2 $J_\textrm{pair}=2$ pairings, leading to blunting or splitting of ZBCPs. At the same time, zero-energy states on BFSs directly contribute to the electron transport at zero bias, which is visible in the absence of ZBCP.

We should comment here that though we have considered BFSs in spin-3/2 $J_{\rm pair}=2$ pairing states, the qualitative behavior of the tunneling spectroscopy generally holds for other multi-band systems. 
Let $\Delta_{\rm inter}$ denote the strength of the TRS-breaking inter-band pair potential. 
The splitting of the BdG spectrum, which is doubly-degenerate for \textit{d}-wave states, is proportional to $(\Delta_{\rm inter}/\mu)^2$, and, thus, the splitting width of the ZEP of the SDOS and ZBCP is also proportional to $(\Delta_{\rm inter}/\mu)^2$. 
In the absence of ZBCP, the charge conductance at zero bias is proportional to the area of BFSs, which scales as $(\Delta_{\rm inter}/\mu)^2$.
\red{Since the factor $(\Delta_{\rm inter}/\mu)^2$ is usually very small, the low-energy bulk LDOS may not be an appropriate amount to experimentally study the contribution of BFSs.
On the other hand, it can be enhanced in materials with very small normal-state Fermi surface, such as the iron-based superconductor \ce{Fe(Se, S)}~\cite{Sato2018Abrupt, Hanaguri2018, Mizukami2023unusual}.}

There are several remaining problems. Recently, the relevance of odd-frequency pairing and BFSs are discussed~\cite{KimAsanoJPSJ,Dutta,HoshinoPRB2021}. 
It is reasonable that the odd-frequency pair amplitude exists even in the bulk state since a gapless superconducting state is realized. 
On the other hand, it is known that an even-frequency superconductor has a nonzero odd-frequency pairing amplitude at an interface or a surface due to the breaking of the translational symmetry~\cite{odd1,odd3,odd3b,tanaka2012}. 
Especially, in the presence of zero energy SABS, the magnitude of the odd frequency pairing at the interface is amplified.   
It is an interesting future work to analyze the obtained results focusing on the symmetry of the Cooper pair including odd-frequency pairings 
and resulting anomalous proximity effect \cite{odd1,Pwaveproximity1,Pwaveproximity2}.

Another possible direction is to study the effect of a magnetic field on the SABSs.
Under an applied magnetic field, the Doppler shift of quasiparticle energy spectra occurs, resulting in the splitting of the ZBCP in spin-singlet \textit{d}-wave superconductors~\cite{Covington1997,Fogelstrom1997}, which can be understood as the consequence of the emergent BFSs due to the TRS-symmetry-breaking external magnetic field~\cite{Setty2020, zhu2020}.
The splitting of the ZBCP under an external magnetic field is also discussed in 3D chiral superconductors and 3D topological insulator/superconductor hybrid systems~\cite{Tamura2017, burset_2015_Superconducting}.
Since the magnetic field is also expected to change the volume of the BFSs in a multi-band system, tunneling spectroscopy under a magnetic field may provide a way to detect the effects of BFSs. 
On the other hand, there is an argument that a large BFS induces instability of superconductivity~\cite{Agterberg2017,Menke2019,Bhattacharya2023}, and whether a large BFS can be realized remains an open question.

\acknowledgments
This work was supported by 
JSPS KAKENHI (Grant Numbers JP19H01824, JP19K14612, JP20H00131,	JP21H01039, 
JP22K03478, and JP23K17668) and JST CREST (Grant No. JPMJCR19T2).

\appendix
\begin{appendix}
    
\section{
Conditions for the appearance of BFSs}
\label{sec:app_Pf}
As a general property, the BdG Hamiltonian possesses particle-hole symmetry:
\begin{align}
    C H_{-\kk} C^{-1}= -H_\kk
\end{align}
where $C=\mathcal{K}\tau_x$
with $\tau_{i=x,y,z}$ being the Pauli matrices in the Nambu space and $\mathcal{K}$ being the complex conjugation operator.
Under the assumption $H_0(\kk)=H_0(-\kk)$,
the Hamiltonian is also symmetric under inversion:
\begin{align}
    P H_{-\kk}P^{-1} = H_\kk,
\end{align}
where $P=\tau_0$
($P=\tau_z$) for even-parity (odd-parity) pairing states. 
Thus, the product of these symmetries satisfies
\begin{align}
    (CP)H_\kk(CP)^{-1} = -H_\kk.
\end{align}
It is generally proved that any $CP$ symmetric Hamiltonian $H_\kk$ can be unitarily transformed into an antisymmetric matrix $\tilde{H}_\kk$, if the square of $CP$ is unity: $(CP)^2=+1$, which is the case for even-parity pairing states~\cite{Agterberg2017}.
To be more concrete, we obtain the antisymmetric matrix, $\tilde{H}_\kk=-\tilde{H}_\kk^{\rm T}$, by 
the unitary transformation $\tilde{H}_\kk = \Omega H_\kk \Omega^\dagger$, where
\begin{align}
    \Omega = \frac{1}{\sqrt{2}}\begin{pmatrix} 1 & 1 \\ -i & i\end{pmatrix}
\end{align}
is a unitary matrix acting on the Nambu space.
It follows that the eigenspectrum of $H_\kk$ is characterized by the Pfaffian of $\tilde{H}_\kk$, ${\rm Pf}(\tilde{H}_\kk)$, which is always real owing to the relation ${\rm Pf}(\tilde{H}_\kk)^2={\rm det}H_\kk>0$ 
for two-band systems.
In general, when the number of bands is even (odd), ${\rm Pf}(\tilde{H}_\kk)$ is always real (purely imaginary), because $H_\kk$ always has pairs of positive and negative eigenvalues due to particle-hole symmetry.
Thus, the Brillouin zone is divided into two regions according to the signs of ${\rm Pf}(\tilde{H}_\kk)$, and zero-energy states appear at the boundary between the two regions. 
In the case of a three-dimensional system, zero-energy states appear on a surface. That is the BFS~\cite{Agterberg2017,Brydon2018}.

The discussion above just says that BFSs {\it may} appear in even-parity pairing states. 
Below, we show that BFSs indeed appear when (i) the intra-band pair potential has nodes, and (ii) the inter-band pairing breaks TRS.
For the even-parity pairing state given by Eq.~\eqref{eq:Delta_2band}, the Pfaffian is calculated as
\begin{align}
    P(\kk)\equiv& \Pf(\tilde{H}_\kk) \nonumber\\
    =&\left|\psi_{\kk,+}\psi_{\kk,-} - \psi_{\kk,I}^2+\dd_\kk\cdot\dd_\kk \right|^2 \nonumber\\
    &+\left(\xi_{\kk,+}\xi_{\kk,-}+\dd_\kk\cdot\dd_\kk^* + |\psi_{\kk,I}|^2\right)^2 \nonumber\\
    &+\xi_{\kk,+}^2|\psi_{\kk,-}|^2+\xi_{\kk,-}^2|\psi_{\kk,+}|^2\nonumber\\
    &-(\dd_\kk\cdot\dd_\kk^* + |\psi_{\kk,I}|^2)^2,
\label{eq:Pf_gen}
\end{align}
where $\xi_{\kk,\pm}=E_{\kk,\pm}-\mu$.
Suppose that the $+$ band has a normal-state Fermi surface, ${\rm FS}_+\equiv \{\kk|\xi_{\kk,+}=0\}$, and its intra-band pair potential $\psi_{\bm k,+}$ has a node along a certain direction.
Along the nodal direction, the Pfaffian~\eqref{eq:Pf_gen} reduces to
\begin{align}
    \left.P(\kk)\right|_{\psi_{\bm k,+}=0}=&\left(\xi_{\kk,+}\xi_{\kk,-}+\dd_\kk\cdot\dd_\kk^* + |\psi_{\kk,I}|^2\right)^2 \nonumber\\
    &-\left|\dd_\kk\times \dd_\kk^*\right|^2-4 \left|{\rm Re}(\psi_{\kk,I}^* \dd_\kk)\right|^2\nonumber\\
    &+\xi_{\kk,+}^2|\psi_{\kk,-}|^2.
\label{eq:Pf_node}
\end{align}
We further assume that the energy scales of the inter-band and intra-band pairings are much smaller than the energy splitting $|E_{\kk,+}-E_{\kk,-}|$ in the vicinity of ${\rm FS}_+$.
Then, there should exist a point $\kk_0$ in the vicinity of ${\rm FS}_+$ such that the first term of the right hand side of Eq.~\eqref{eq:Pf_node} disappears.
The Pfaffian at $\kk=\kk_0$ is then given by
\begin{align}
    \left.P(\kk)\right|_{\kk=\kk_0}=&-\left|\dd_\kk\times \dd_\kk^*\right|^2-4 \left|{\rm Re}(\psi_{\kk,I}^* \dd_\kk)\right|^2\nonumber\\
    &+\frac{|\psi_{\kk,-}|^2(\dd_\kk\cdot \dd_\kk^*+|\psi_{\kk,I}|^2)^2}{\xi_{\kk,-}^2}.
\label{eq:Pf_min}
\end{align}
From the assumption that $|\xi_{\kk,-}|\simeq|E_{\kk,+}-E_{\kk,-}|$ is much larger than the pair potentials,
the last term of Eq.~\eqref{eq:Pf_min} is negligible,
and the right-hand-side of Eq.~\eqref{eq:Pf_min} becomes negative if the inter-band pairing breaks TRS [see Eq.~\eqref{eq:spin_polarization}].
On the other hand, far from the normal state Fermi surfaces of both $+$ and $-$ bands,
the Pfaffian~\eqref{eq:Pf_gen} becomes positive because of the $(\xi_{\kk,+}\xi_{\kk,-})^2$ term.
Thus, the sign of the Pfaffian changes in the Brillouin zone and BFSs appear.

\section{\texorpdfstring{$\bm{J=3/2}$}{} representation of the rank-2 tensor}
\label{sec:app_matrix}
Using the $4\times 4$ matrix form of the angular momentum operator $\bm J$,
\begin{align}
J_x&=\begin{pmatrix} 0 & \sqrt{3}/2 & 0 & 0 \\
\sqrt{3}/2 & 0 & 1 & 0 \\
0 & 1 & 0 &\sqrt{3}/2\\
0 & 0 & \sqrt{3}/2 & 0 \end{pmatrix},\\
J_y&=\begin{pmatrix} 0 & -i\sqrt{3}/2 & 0 & 0 \\
i\sqrt{3}/2 & 0 & -i & 0 \\
0 & i & 0& -i\sqrt{3}/2\\
0 & 0 & i\sqrt{3}/2 & 0 \end{pmatrix},\\
J_z&=\begin{pmatrix}3/2 & 0 & 0 & 0 \\
0&1/2 & 0 & 0 \\
0&0&-1/2&0\\
0&0&0&-3/2\end{pmatrix},
\end{align}
the the rank-2 tensor $\mathscr{Y}_j(\bm J)\ (j=1,2,3,4,5)$ is given by
\begin{align}
\mathscr{Y}_1(\bm J)&=\begin{pmatrix} 0 & 0 & -i & 0 \\ 0 & 0 & 0 & -i \\ i & 0 & 0& 0\\ 0 & i & 0 & 0\end{pmatrix}=\tau_y\otimes\sigma_0,\\ 
\mathscr{Y}_2(\bm J)&=\begin{pmatrix} 0 &  -i & 0&0 \\ i&0 & 0 & 0 \\  0 & 0& 0&i\\ 0 &0& -i & 0\end{pmatrix}=\tau_z\otimes\sigma_y,\\
\mathscr{Y}_3(\bm J)&=\begin{pmatrix} 0 &  1 & 0&0 \\ 1&0 & 0 & 0 \\  0 & 0& 0&-1\\ 0 &0& -1 & 0\end{pmatrix}=\tau_z\otimes\sigma_x,\\ 
\mathscr{Y}_4(\bm J)&=\begin{pmatrix} 0 & 0 & 1 & 0 \\ 0 & 0 & 0 & 1 \\ 1 & 0 & 0& 0\\ 0 & 1 & 0 & 0\end{pmatrix}=\tau_x\otimes\sigma_0,\\
\mathscr{Y}_5(\bm J)&=\begin{pmatrix} 1 & 0 & 0 & 0 \\ 0 & -1 & 0 & 0 \\ 0 & 0 & -1& 0\\ 0 & 0 & 0 & 1\end{pmatrix}=\tau_z\otimes\sigma_z.
\end{align}

\section{Tight-binding Hamiltonian}
\label{sec:app_tb}
We summarize the matrix in the tight-binding model along the $z$ direction.  
\subsection{spin-1/2 system}
The hopping matrix $t(\kk_\perp)$ and on-site potential matrix $u(\kk_\perp)$ in the normal-state Hamiltonian is derived from $H_0(\kk)=\xi_{\kk,+}\otimes\mathbbm{1}_2$ as
\begin{align}
    \bm t(\bm k_\perp)&=\frac{\hbar^2}{2M_+a^2}\mathbbm{1}_2,\\
    \bm u(\bm k_\perp)&=\left[\frac{\hbar^2}{2M_+}\left(k_x^2+k_y^2+\frac{2}{a^2}\right)-\mu\right]\mathbbm{1}_2.
\end{align}
The pair potential with the \textit{d}-wave symmetry given by Eq.~\eqref{eq:3states_spin1/2} includes both the on-site and off-site pairing terms. For each state in Eq.~\eqref{eq:3states_spin1/2}, the pair potentials are given as follows:
\begin{align}
&T_{2g}(0,i,1):\nonumber\\
&\hspace*{10mm}\Delta_{\rm on}(\kk_\perp)=0,\\
&\hspace*{10mm}\Delta_{\rm off}(\kk_\perp)=\frac{\sqrt{3}\Delta_0}{2a}(-ik_x+k_y) i\sigma_y,\\
&T_{2g}(1,\omega,\omega^2):\nonumber\\
&\hspace*{10mm}\Delta_{\rm on}(\kk_\perp)=\frac{\sqrt{3}\Delta_0}{2}k_x k_y i\sigma_y,\\
&\hspace*{10mm}\Delta_{\rm off}(\kk_\perp)=-i\frac{\sqrt{3}\Delta_0}{2a}(\omega k_x + \omega^2k_x) i\sigma_y,\\
&E_g(1,i):\nonumber\\
&\hspace*{10mm}\Delta_{\rm on}(\kk_\perp)=\frac{\Delta_0e^{i\pi/6}}{2}\left(k_x^2+\omega k_y^2+\frac{2\omega^2}{a^2}\right) i\sigma_y,\\
&\hspace*{10mm}\Delta_{\rm off}(\kk_\perp)=-\frac{\omega^2\Delta_0e^{i\pi/6}}{a^2} i\sigma_y.
\end{align}

\subsection{spin-3/2 system}
The hopping matrix $t(\kk_\perp)$ and on-site potential matrix $u(\kk_\perp)$ in the normal-state Hamiltonian is derived from Eq.~\eqref{eq:H0_Oh} with $m_1=m_2$ as
\begin{align}
    \bm t(\bm k_\perp)=&\frac{\hbar^2}{2m_0a^2}\mathbbm{1}_4    +\frac{\hbar^2}{2m_1a^2}\mathscr{Y}_5(\bm J)\nonumber\\
    &-i\frac{\sqrt{3}\hbar^2}{2m_1a}\left[k_x\mathscr{Y}_1(\bm J)+k_y\mathscr{Y}_2(\bm J)\right],\\
    \bm u(\bm k_\perp)=&\left[\frac{\hbar^2}{2m_0}\left(k_x^2+k_y^2+\frac{2}{a^2}\right)-\mu\right]\mathbbm{1}_4\nonumber\\
    &+\frac{\sqrt{3}\hbar^2}{2m_1}k_xk_y\mathscr{Y}_1(\bm J)\nonumber\\
    &+\frac{\sqrt{3}\hbar^2}{4m_1}\left(k_x^2-k_y^2\right)\mathscr{Y}_4(\bm J)\nonumber\\
    &+\frac{\hbar^2}{4m_1}\left(\frac{4}{a^2}-k_x^2-k_y^2\right)\mathscr{Y}_5(\bm J).
\end{align}
We consider $\kk$-independent $J_{\rm pair}=2$ pair potentials, which leads to $\Delta_{\rm off}=0.$
The on-site pair potentials corresponding to Eqs.~\eqref{eq:3states_spin3/2_a}-\eqref{eq:3states_spin3/2_c} are given by
\begin{align}
&T_{2g}(0,i,1):\nonumber\\
&\hspace*{10mm}\Delta_{\rm on}=\Delta_0\left[i\mathscr{Y}_2(\bm J)+\mathscr{Y}_3(\bm J)\right],\\
&T_{2g}(1,\omega,\omega^2):\nonumber\\
&\hspace*{10mm}\Delta_{\rm on}=\Delta_0\left[\mathscr{Y}_1(\bm J)+\omega\mathscr{Y}_2(\bm J)+\omega^2\mathscr{Y}_3(\bm J)\right],\\
&E_g(1,i):\nonumber\\
&\hspace*{10mm}\Delta_{\rm on}=\Delta_0\left[\mathscr{Y}_4(\bm J)+i\mathscr{Y}_5(\bm J)\right].
\end{align}

\section{Generalized Lee-Fisher formula in Nambu space}
\label{sec:app_LF}
We derive the Lee-Fisher formula~\cite{Lee1981} generalized for the case with internal degrees of freedom in Nambu space.
We consider a normal-state Hamiltonian with generic spin-dependent hopping terms
\begin{align}
    \mathcal{H}_{\rm hop}&=-\sum_{l,\kk_\perp} \sum_{\lambda,\lambda'} \left[  f_{l+1,\kk_\perp,\lambda}^\dagger t_{\lambda\lambda'}(\kk_\perp) f_{l,\kk_\perp,\lambda'} + \textrm{H. c.}\right],\\
    &=-\sum_{l,\kk_\perp}\left[ (\bm f_{l+1,\kk_\perp}^\dagger)^{\rm T} \bm t(\kk_\perp) \bm f_{l,\kk_\perp} + \textrm{H. c.}\right],
\end{align}
where $t_{\lambda\lambda'}(\kk_\perp)$ is an element of the hopping matrix with $\lambda$ and $\lambda'$ being the indices for the internal degrees of freedom.
The second line is just rewriting the first line using the matrix forms.
We can rewrite $\mathcal{H}_{\rm hop}$ using the Nambu representation as
\begin{widetext}
\begin{align}
    \mathcal{H}_{\rm hop}=-\frac{1}{2}\sum_{l,\kk_\perp}\begin{pmatrix} (\bm f_{l+1,\kk_\perp}^\dagger)^{\rm T} & (\bm f_{l+1,-\kk_\perp})^{\rm T} & (\bm f_{l,\kk_\perp}^\dagger)^{\rm T}  & (\bm f_{l,-\kk_\perp})^{\rm T}\end{pmatrix}
    \begin{pmatrix} \bm 0 & \tau_3\bm T(\kk_\perp) \\[2mm] \bm \tau_3\bm T^\dagger(\kk_\perp) & \bm 0 \end{pmatrix}
    \begin{pmatrix} \bm f_{l+1,\kk_\perp}  \\ \bm f_{l+1,-\kk_\perp}^\dagger \\ \bm f_{l,\kk_\perp} \\ \bm f_{l,-\kk_\perp}^\dagger\end{pmatrix},
\label{eq:H_hop_append}
\end{align}
where $\bm T(\kk_\perp)$ is given by Eq.~\eqref{eq:Tkk}.
Here, $\tau_3$ in front of $\bm T(\bm k_\perp)$ and $\bm T^\dagger(\bm k_\perp)$ comes from the Fermi statistics.
The current operator $J$ is defined from the equation of continuity:
\begin{align}
-|e|\frac{\partial}{\partial t} \sum_{\kk_\perp}\left(\bm{f}_{l,\kk_\perp}^{\dagger}\right)^{\rm T} \bm{f}_{l,\kk_\perp} + \frac{J(l)-J(l-1)}{a} = 0,
\end{align}
where $e<0$ is the elementary charge and $a$ is the lattice constant.
Considering the fact that the current comes from the hopping Hamiltonian, 
the current operator at site $l$ is given by
\begin{align}
    J(l)=&
    \frac{|e|a}{i\hbar}\sum_{\kk_\perp}\left[(\bm f_{l+1,\kk_\perp}^\dagger)^{\rm T} \bm t(\kk_\perp) \bm f_{l,\kk_\perp}-\textrm{H. c.}\right]\nonumber\\
    =&\frac{|e|a}{i2\hbar}\sum_{\kk_\perp} \begin{pmatrix} (\bm f_{l+1,\kk_\perp}^\dagger)^{\rm T} & (\bm f_{l+1,-\kk_\perp})^{\rm T} & (\bm f_{l,\kk_\perp}^\dagger)^{\rm T}  & (\bm f_{l,-\kk_\perp})^{\rm T}\end{pmatrix}
    \begin{pmatrix} \bm 0 & \bm T(\kk_\perp) \\[2mm] -\bm T^\dagger(\kk_\perp) & \bm 0 \end{pmatrix}
        \begin{pmatrix} \bm f_{l+1,\kk_\perp}  \\ \bm f_{l+1,-\kk_\perp}^\dagger \\ \bm f_{l,\kk_\perp} \\ \bm f_{l,-\kk_\perp}^\dagger\end{pmatrix},
\end{align}
\end{widetext}
which indicates that $\bm T(\kk_\perp)$ is the hopping matrix for the current operator in the Nambu description.
Using this $\bm T(\kk_\perp)$ and following the procedure to derive the Lee-Fisher formula~\cite{Lee1981}, see, e.g., Appendix 1 of Ref.~\cite{Inoue_text} for details, we obtain Eq.~\eqref{eq:Lee-Fisher}.

\end{appendix}

%

\end{document}

%% file: dispersion.tex
\begin{tikzpicture}[scale=0.7] 
\draw[-triangle 45,name path=E] (0,-3) -- (0,3) node[right,scale=2]{$E$}; 
\draw[-triangle 45,name path=K] (-4,.5) -- (4,.5) node[below right,scale=2]{$\bm{k}$}; 

\draw[dashed] (0,-1) -- (-1,-1) node[left, scale=1.5]{$-\mu$}; 

\draw[line width=.8mm,teal,name path=Em, line cap=round] (-2,-2.5) parabola bend (0,-1.1) (2,-2.5) node[right,scale=2]{$\xi_{\bm{k},-}$}; 
\draw[line width=.8mm,brown,name path=Ep, line cap=round] (-2.5,2.5) parabola bend (0,-1) (2.5,2.5) node[below right,scale=2] {$\xi_{\bm{k},+}$}; 

\path[name intersections={of= Ep and K}]; 
\path node[below left, scale=1.5] at (intersection-1) {$-k_\mathrm{F}$}; 
\path node[below right, scale=1.5] at (intersection-2) {$k_\mathrm{F}$}; 

\path[name intersections={of= E and K}]; 
\path node[below left,scale=1.2] at (intersection-1){$\bm{0}$}; 

\end{tikzpicture} 